\documentclass[preprint,preprintnumbers,amsmath,amssymb,nofootinbib]{revtex4}

\usepackage{graphicx}
\usepackage{amsmath}
\usepackage{amssymb}
\usepackage{graphicx}
\usepackage{dsfont}
\usepackage{overpic}
\usepackage{color}
\usepackage{slashed}
\usepackage{subfigure}
\usepackage{graphicx,textcomp}
\usepackage{natbib}

\allowdisplaybreaks

\newcommand{\bl}{\bar{\lambda}}

\begin{document}
\title{Pion correlation functions in position space from chiral perturbation theory with resonance exchange}
\author{Peter C.~Bruns}
\affiliation{Institut f\"ur Theoretische Physik, Universit\"at Regensburg, D-93040 Regensburg, Germany}
\date{\today}
\begin{abstract}
Explicit expressions for pion correlators are derived in position-space, employing Chiral Perturbation Theory (ChPT). Resonance exchange contributions are included to test the range of applicability of the leading-order ChPT expressions. The obtained results will be useful for a forthcoming Lattice-QCD study of double parton distributions in the pion.
\end{abstract}
\maketitle
\section{Generalities}

\begin{figure}[h]
\centering
\includegraphics[width=0.50\textwidth]{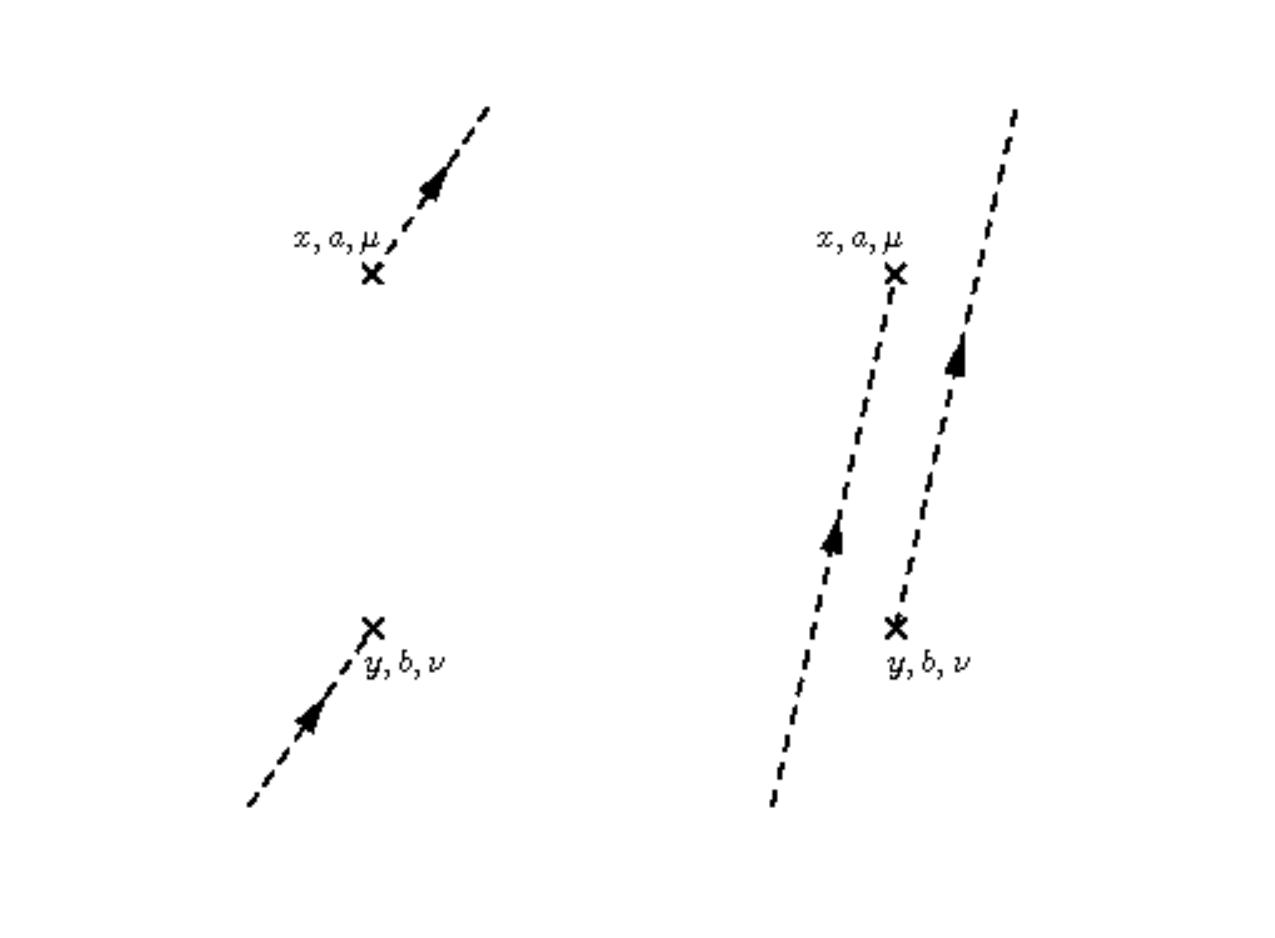}
\caption{Two disconnected Feynman graphs contributing to {\bf PP,\,AA} at leading chiral order. The dashed lines represent the pions. In this work, only the results from fully connected graphs are given explicitly.}
\label{fig:fmgraphs3}
\end{figure}

We consider matrix elements of the form 
\begin{eqnarray*}
C_{PP}^{00}(x,y) &:=& \langle \pi(p')|TP^{0}\left(x\right)P^{0}\left(y\right)|\pi(p)\rangle\,,\quad C_{SS}^{00}(x,y) := \langle \pi(p')|TS^{0}\left(x\right)S^{0}\left(y\right)|\pi(p)\rangle\,,\\
C_{PP}^{ab}(x,y) &:=& \langle \pi(p')|TP^{a}\left(x\right)P^{b}\left(y\right)|\pi(p)\rangle\,,\quad C_{SS}^{ab}(x,y) := \langle \pi(p')|TS^{a}\left(x\right)S^{b}\left(y\right)|\pi(p)\rangle\,,\\
C_{AA,\mu\nu}^{ab}(x,y) &:=& \langle \pi(p')|TA^{a}_{\mu}\left(x\right)A^{b}_{\nu}\left(y\right)|\pi(p)\rangle\,,\quad C_{VV,\mu\nu}^{ab}(x,y) := \langle \pi(p')|TV^{a}_{\mu}\left(x\right)V^{b}_{\nu}\left(y\right)|\pi(p)\rangle\,,
\end{eqnarray*}
with position four-vectors $x^{\mu}$, $y^{\mu}$ and currents 
\begin{eqnarray}
P^{a}(x) &:=& i\bar{q}(x)\gamma_{5}\tau^{a}q(x)\,,\quad P^{0}(x):= i\bar{q}(x)\gamma_{5}q(x)\,,\quad S^{a}(x):= \bar{q}(x)\tau^{a}q(x)\,,\nonumber \\
A^{a}_{\mu}(x) &:=& \bar{q}(x)\gamma_{\mu}\gamma_{5}\frac{\tau^{a}}{2}q(x)\,,\quad V^{a}_{\mu}(x) := \bar{q}(x)\gamma_{\mu}\frac{\tau^{a}}{2}q(x)\,,\quad S^{0}(x):= \bar{q}(x)q(x)\,,\label{eq:opdefs}
\end{eqnarray}
where $\tau^{a}$, $a=1,2,3$, are the Pauli matrices, and we have suppressed here the channel (particle species) or isospin indices of the pions.
Here we are interested in the non-trivial dependence of these matrix elements on the distance $x-y$. More specifically, we are interested in those contributions which stem from an exchange of one or more particles between the vertices at $x$ and $y$ which give the most important contributions for large distances $|\vec{x}-\vec{y}|\gtrsim\frac{1}{M_{\pi}}$. We will therefore disregard contact-term contributions $\sim\delta^{3}(\vec{x}-\vec{y})$, as well as disconnected graphs, e.~g., graphs where the external pions are not connected to the vertices at $x$ and $y$. The latter are of the form
\begin{equation} 
C_{\mathcal{O}\mathcal{O}}^{\mathrm{spect.}}(x-y) = 2E_{\mathbf{p}}(2\pi)^{3}\delta^{3}(\mathbf{p'}-\mathbf{p})\langle 0|T\mathcal{O}\left(x\right)\mathcal{O}\left(y\right)|0\rangle\,,
\end{equation}
where $E_{\mathbf{p}}=\sqrt{|\mathbf{p}|^2+M_{\pi}^{2}}\,$. The ChPT results for the vacuum expectation values on the r.h.s. (at the one-loop level) can e.~g. be found in Sec.~12 and 13 in \cite{Gasser:1983yg}. Moreover, disconnected graphs like the ones in Fig.~\ref{fig:fmgraphs3} do not lead to a non-trivial dependence on the distance, and will not be considered here. The contributions from fully connected graphs, on which we will focus in this work, are subsumed in $C_{\mathcal{O}\mathcal{O}}^{\mathrm{conn.}}$. We will compute the dominant long-range part of this quantity here, in the framework of ChPT \cite{Gasser:1983yg,Weinberg:1978kz}, following the general methods spelled out e.~g. in the textbook \cite{Weinberg:1995mt} (see Chapter 6 and Eqs.~(10.4.19,20) therein, and also Chapter 9 of \cite{Itzykson:1980rh}), which are based on Green functions, path integrals and the effective action \cite{Schwinger:1951ex,Schwinger:1951xk,Feynman:1948ur,Lehmann:1954rq,Coleman:1973jx}. We will also test the range of applicability of those ChPT formulae by a comparison to some specific higher-order contributions generated by tree-level resonance exchange graphs. The necessary Feyman rules and Fourier integrals are collected in App.~\ref{app:frules} and \ref{app:integrale}. In App.~\ref{app:softpions}, we include the demonstration of some general relations between the matrix elements in the limit of vanishing pion masses and momenta, which are well-known in the literature as ``soft-pion'' theorems. The results will be relevant for a forthcoming study of double parton distributions in the pion \cite{RQCD}, employing Lattice QCD (see also \cite{Burkardt:1994pw} for an earlier lattice study).

\subsection{General structure of the contributions}

Schematically, the contributions from the connected graphs are of the form 
\begin{equation}
G(p,p',q_{\mathrm{ext}},q'_{\mathrm{ext}})=\int\frac{d^{d}q_{i1}}{(2\pi)^{d}}\int\frac{d^{d}q_{i2}}{(2\pi)^{d}}\ldots M\left(p,p',q_{\mathrm{ext}},q'_{\mathrm{ext}},q_{i1},q_{i2},\ldots\right)\prod_{V}(2\pi)^{d}\delta^{d}(q_{V})\,
\end{equation}
in $d$-dimensional momentum space.
The delta functions assure four-momentum conservation at every vertex $V$. $M$ is the remainder of the amplitude, where all delta functions have been extracted. Integration over all internal four-momenta $q_{i1},q_{i2}\ldots$ leaves us with only one delta function $\delta^{d}(p'+q'_{\mathrm{ext}}-p-q_{\mathrm{ext}})$, which expresses the overall momentum conservation. Here $q_{\mathrm{ext}},q'_{\mathrm{ext}}$ label the four-momenta running in or out of the diagram at $y$ and $x$, respectively, transferred by the external source fields $v,a,s,p$ (see below). The amplitude in position space is obtained by Fourier transformation (see e.~g.~\cite{Itzykson:1980rh}, Eq.~(6-20)),
\begin{equation}\label{eq:Corr}
C_{\cdots}(x,y) = \int\frac{d^{d}q'_{\mathrm{ext}}}{(2\pi)^{d}}\int\frac{d^{d}q_{\mathrm{ext}}}{(2\pi)^{d}}e^{-iq'_{\mathrm{ext}}x}e^{iq_{\mathrm{ext}}y}(2\pi)^{d}\delta^{d}(p'+q'_{\mathrm{ext}}-p-q_{\mathrm{ext}}) \bar{M}\left(p,p',q_{\mathrm{ext}},q'_{\mathrm{ext}}\right)\,.
\end{equation}
Here $\bar{M}$ directly results from the integration over internal four-momenta in $M$. In the specific framework of ChPT applied here, the generating functional of all QCD correlators is evaluated by means of a path integral involving an {\em effective}\, low-energy Lagrangian $\mathcal{L}_{\mathrm{eff}}(U,v,a,s,p\ldots)$ (see \cite{Gasser:1983yg}, and Eqs.~(1) and (2) of \cite{Ecker:1989yg}),
\begin{eqnarray}
e^{iZ\lbrack v,a,s,p\rbrack} &=& \langle 0|\,T\exp\biggl(i\int d^{4}x\,\bar{q}\lbrack\gamma_{\mu}(v^{\mu}+\gamma_{5}a^{\mu})-(s-ip\gamma_{5})\rbrack q\biggr)|0\rangle \nonumber \\
 &=&  \int\lbrack dU\rbrack \exp\biggl(i\int d^{4}x\,\mathcal{L}_{\mathrm{eff}}(U,v,a,s,p\ldots)\biggr)\,.\label{eq:Zfunctional}
\end{eqnarray}
Formally, all QCD Green functions can be obtained by taking functional derivatives of the generating functional w.r.t. the external source fields $s,p,v^{\mu},a^{\mu},\ldots$ (it is possible to extend the analysis to higher-rank tensor fields to analyze matrix elements of more complicated operators (see e.~g.~\cite{Donoghue:1991qv,Arndt:2001ye,Kivel:2002ia,Diehl:2005rn})). The matrix field $U$ collects the pion (Goldstone boson) fields in a convenient way (see below). The effective Lagrangian has to be invariant under {\em local}\, chiral transformations of the Goldstone boson and source fields, and shares all other symmetries of $\mathcal{L}_{\mathrm{QCD}}$. A formal proof that low-energy QCD can indeed be analyzed in this way has been given by Leutwyler \cite{Leutwyler:1993iq}.
The effective Lagrangian and the perturbation series are ordered by a low-energy power counting scheme, counting suppression powers of Goldstone boson momenta and masses (or quark masses). For details and further references, we refer to \cite{Gasser:1983yg,Leutwyler:1993iq,Scherer:2002tk}. Since the Feynman rules are read off from $i\mathcal{L}_{\mathrm{eff}}$, we note from a comparison of the l.h.s. and the r.h.s. of Eq.~(\ref{eq:Zfunctional}) that we must multiply our graphs with a phase factor $(-i)^n(+i)^m$ to obtain the result for a correlator involving $n$ operators $P(x),V(x)$ or $A(x)$ and $m$ operators $S(x)$. This is because the factors of $i$, which we include in our Feynman rules also for vertices with external sources, are cancelled when taking functional derivatives $\delta/\delta ip$, $-\delta/\delta is\,$\ldots of Eq.~(\ref{eq:Zfunctional}).\\
At leading chiral order, the effective Lagrangian is given by (see \cite{Gasser:1983yg,Weinberg:1978kz})
\begin{equation}\label{eq:L2M}
\mathcal{L}_{M}^{(2)} = \frac{F^{2}}{4}\langle\nabla_{\mu}U^{\dagger}\nabla^{\mu}U\rangle+\frac{F^{2}}{4}\langle \chi U^{\dagger}+U\chi^{\dagger}\rangle\,,
\end{equation}
with $\chi=2B(s+ip)$, $s=\mathcal{M}+\delta s$, where $\mathcal{M}$ is the quark mass matrix, and $\delta s$ the remaining part of $s$. The brackets $\langle\ldots\rangle$ denote the flavor (or isospin) trace, $F$ is the pion decay constant in the chiral limit, and
\begin{displaymath}
\nabla_{\mu}U = \partial_{\mu}U-i(v_{\mu}+a_{\mu})U+iU(v_{\mu}-a_{\mu})\,.
\end{displaymath}
Here $U=\exp(i\sqrt{2}\phi/F)$ with $\phi = \phi^{j}\bar{\lambda}^{j}$, where $j$ is a channel (particle species) index which labels the specific pion, and $\bar{\lambda}$ are the pertaining channel matrices. We write out $\phi$ as
\begin{displaymath}
\phi = \pi^{0}\bar{\lambda}^{\pi^{0}}+\pi^{+}\bar{\lambda}^{\pi^{+}}+\pi^{-}\bar{\lambda}^{\pi^{-}}\,,
\end{displaymath}
see also App.~\ref{app:frules}.
Let us consider an example with just one operator insertion: From the vertex rule (\ref{eq:leadingpvertex}), we find 
\begin{eqnarray}\label{eq:Gpi}
\langle 0|P^{a}(x)|\pi^{j}(p)\rangle &=& \int\frac{d^{d}q_{\mathrm{ext}}}{(2\pi)^{d}}e^{iq_{\mathrm{ext}}x}(2\pi)^{d}\delta^{d}(p+q_{\mathrm{ext}})\left(\sqrt{2}BF\langle\bar{\lambda}^{j}\tau^{a}\rangle + \mathcal{O}(p^2)\right) \nonumber \\ &=& G_{\pi}e^{-ipx}\frac{\langle\bar{\lambda}^{j}\tau^{a}\rangle}{\sqrt{2}}\,,\quad G_{\pi}=2BF + \mathcal{O}(p^2)\,,
\end{eqnarray}
while the matrix element for the pseudoscalar isosinglet current vanishes at leading order,
\begin{equation}\label{eq:nopi}
\langle 0|P^{0}(x)|\pi^{0}(p)\rangle =  \tilde{G}_{\pi}e^{-ipx}\,,\quad \tilde{G}_{\pi} = 0 + \mathcal{O}(p^2)\,.
\end{equation}
With the same method, and Eq.~(\ref{eq:vertexspipi}) we also find (see e.~g. Eq.~(15.9) in \cite{Gasser:1983yg})
\begin{equation}\label{eq:scalff}
\langle\pi^{k}|\bar{q}q|\pi^{j}\rangle = i\left(-2iB\langle \bar{\lambda}^{j}\bar{\lambda}^{k\dagger}\rangle\right)+\mathcal{O}(p^2) = 2B\delta^{jk}+\mathcal{O}(p^2)\,.
\end{equation}
The pion propagator in momentum space is also easily derived from $\mathcal{L}_{M}^{(2)}$,
\begin{equation}\label{eq:piprop}
i\Delta_{\pi}(q):=\frac{i}{q^2-M_{\pi}^{2}}\,,
\end{equation}
for a four-momentum $q$ of the propagating pion.

\vspace{1cm}

\subsection{Isospin symmetry}

For isoscalar operators, marked with the index $0$, isospin symmetry (which we will assume in the following) requires that the matrix representation of the operator is diagonal in the (cartesian) isospin indizes $c,d=1,2,3$ of the pions,
\begin{equation}
\langle \pi^{d}|T\mathcal{O}^{0}\left(x\right)\mathcal{O}^{0}\left(y\right)|\pi^{c}\rangle = \delta^{cd}C_{\mathcal{O}\mathcal{O}}^{00}(x,y)\,,
\end{equation}
with $C_{\mathcal{O}\mathcal{O}}^{00}(x,y)$ independent of the individual isospin components (a simple consequence of the Wigner-Eckart theorem).
For isovector operators, the symmetry requirement is less trivial: Here there are three independent coefficient functions. One can e.~g. decompose
\begin{eqnarray}
\langle \pi^{d}|T\mathcal{O}^{a}\left(x\right)\mathcal{O}^{b}\left(y\right)|\pi^{c}\rangle &=& \left(\delta^{ab}\delta^{cd}+\delta^{ac}\delta^{bd}+\delta^{ad}\delta^{bc}\right)C_{1}(x,y) + \left(\delta^{ac}\delta^{bd}+\delta^{ad}\delta^{bc}\right)C_{2}(x,y) \nonumber \\ &+& \left(\delta^{ac}\delta^{bd}-\delta^{ad}\delta^{bc}\right)C_{3}(x,y)\,.
\end{eqnarray}
Obviously, the function $C_{3}(x,y)$ has to be antisymmetric under $x\leftrightarrow y$, whereas $C_{1}(x,y)$, $C_{2}(x,y)$ are symmetric under this operation.
As an example, we give the representation of some specific $PP$ isovector matrix elements,
\begin{eqnarray}
\langle \pi^{+}(p')|TP^{3}\left(x\right)P^{3}\left(y\right)|\pi^{+}(p)\rangle &=& C_{1}(x,y)\,,\\
\langle \pi^{0}(p')|TP^{3}\left(x\right)P^{3}\left(y\right)|\pi^{0}(p)\rangle &=& 3C_{1}(x,y) + 2C_{2}(x,y)\,,
\end{eqnarray}
and also for some other isospin structures,
\begin{eqnarray}
\langle\pi^{+}(p')|\bar{u}(x)i\gamma_{5}d(x)\bar{u}(y)i\gamma_{5}d(y)|\pi^{-}(p)\rangle &=& C_{1}(x,y) + C_{2}(x,y)\,,\\
\langle\pi^{+}(p')|\bar{q}(x)i\gamma_{5}\tau^{3}q(x)\bar{u}(y)i\gamma_{5}d(y)|\pi^{0}(p)\rangle &=& \frac{1}{\sqrt{2}}(C_{1}(x,y) + C_{2}(x,y)+C_{3}(x,y))\,,\\
\langle\pi^{+}(p')|\bar{u}(x)i\gamma_{5}d(x)\bar{d}(y)i\gamma_{5}u(y)|\pi^{+}(p)\rangle &=& C_{1}(x,y)+\frac{1}{2}(C_{2}(x,y)-C_{3}(x,y))\,.
\end{eqnarray}
This representation is generally valid assuming isospin symmetry. Note that an additional factor of 4 appears on the r.h.s. of the previous equations if the pseudoscalar operators are replaced by their vector or axial-vector counterparts, because factors of $\frac{1}{2}$ are included in the operator definitions in Eq.~(\ref{eq:opdefs}). The mixed isoscalar-isovector case,
\begin{equation}
\langle \pi^{d}|T\mathcal{O}^{0}\left(x\right)\mathcal{O}^{a}\left(y\right)|\pi^{c}\rangle = i\epsilon^{dac}C_{\mathcal{O}\mathcal{O}}^{\mathrm{mix}}(x,y)\,,
\end{equation}
will not be considered here, as most of the pion-exchange contributions to such correlators vanish at the order we are working.

\newpage

\section{Pion exchange contributions}

\begin{figure}[h]
\centering
\includegraphics[width=0.62\textwidth]{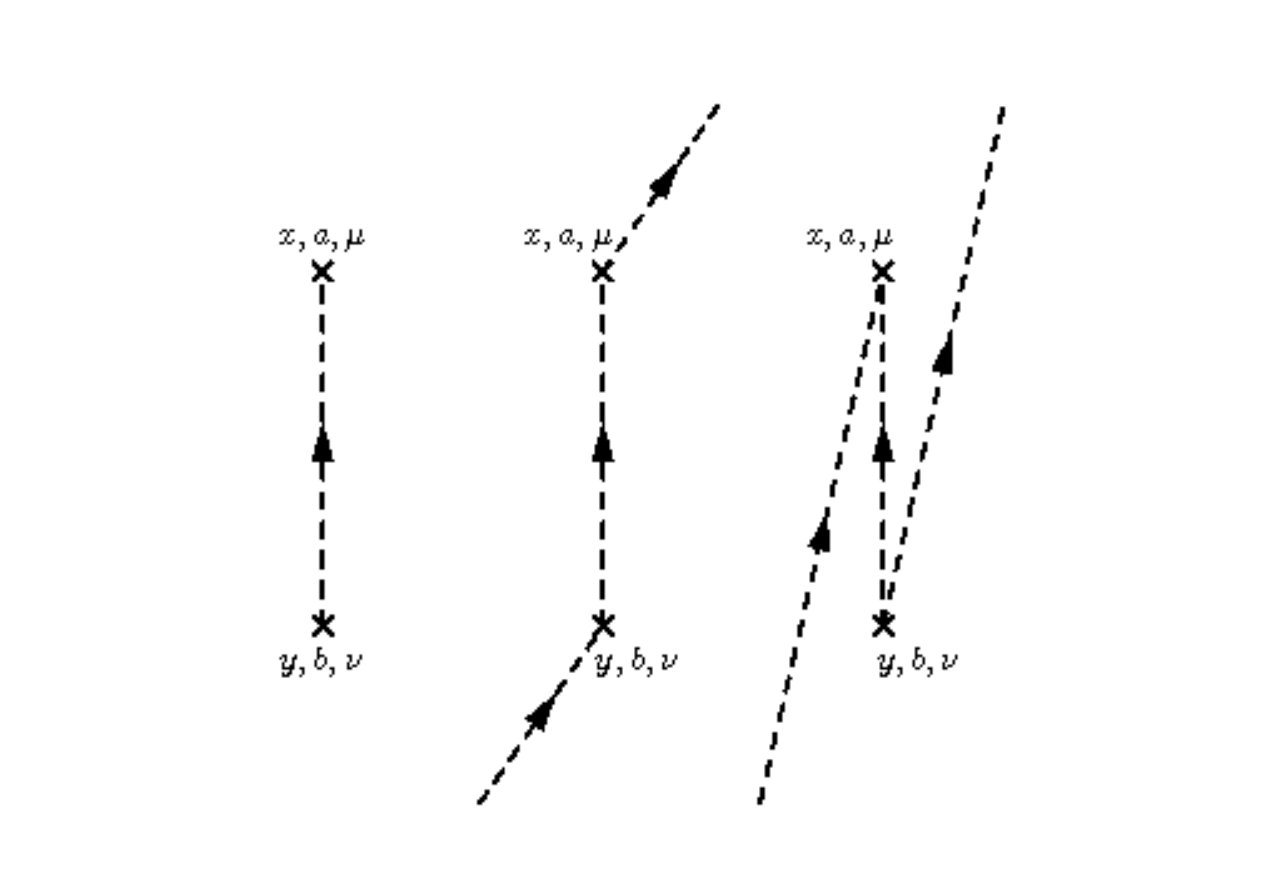}
\caption{Feynman graphs contributing to the vac. exp. value of {\bf PP,\,AA} (first) and to the pion correlators of {\bf SS,\,VV} (second and third graph, ``Z''-topology) at leading chiral order. The crosses stand for the operator insertions at $x$ and $y$.}
\label{fig:fmgraphs1}
\end{figure}

\begin{figure}[h]
\centering
\includegraphics[width=0.62\textwidth]{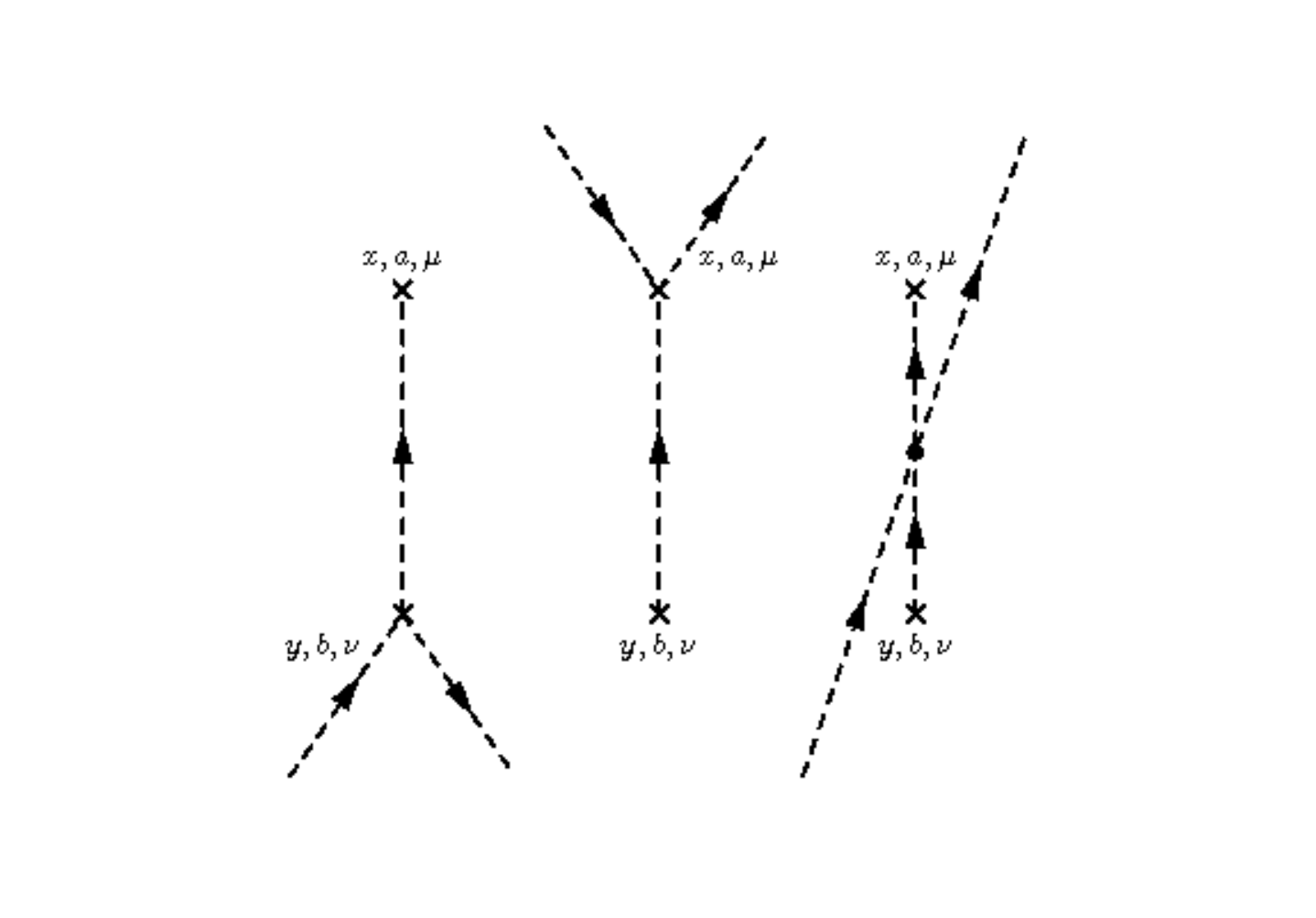}
\caption{Graphs contributing to the pion correlators of {\bf PP,\,AA} at leading chiral order (``T''-and ``X''-topology).}
\label{fig:fmgraphs2}
\end{figure}

Taking into account the Feynman graphs of Figs.~\ref{fig:fmgraphs1} and \ref{fig:fmgraphs2}, tree-level ChPT at leading order gives the following results for the functions $C^{00}_{\cdots}$ and $C_{1-3}^{\,\cdots}\,$: \\
In the case of isovector pseudoscalar operators (indicated by the superscript $PP$), we find
\begin{eqnarray}
C_{1}^{PP} &=& 4B^{2}e^{i\frac{\Delta}{2}(x+y)}\left(M_{\pi}^{2}-\Delta^2\right)I_{\pi\pi}^{\Delta}(x-y)\,,\label{eq:resC1PP}\\
C_{2}^{PP} &=& -4B^{2}e^{i\frac{\Delta}{2}(x+y)}\left(\cos\left(\frac{\Delta}{2}(x-y)\right)I_{\pi}(x-y) + \left(2M_{\pi}^{2}-\frac{3}{2}\Delta^2\right)I_{\pi\pi}^{\Delta}(x-y)\right),\label{eq:resC2PP}\\
C_{3}^{PP} &=& 8B^{2}e^{i\frac{\Delta}{2}(x+y)}\bar{p}\cdot(x-y)I_{\pi\pi}^{\Delta(1)}(x-y)\,.\label{eq:resC3PP}
\end{eqnarray}
We use the notation $\Delta=p'-p$, $\bar{p}=\frac{1}{2}\left(p+p'\right)$, where $p$ ($p'$) is the four-momentum of the incoming (outgoing) pion.
The Fourier integrals $I_{\pi}\equiv I_{M}(M\rightarrow M_{\pi})$, $I_{\pi\pi}^{\Delta}\equiv I_{MM}^{\Delta}(M\rightarrow M_{\pi})$ etc. can all be found in App.~\ref{app:integrale}.
The function $C_{3}^{PP}(x,y)$ vanishes if $\bar{p}\cdot(x-y)=0$\,. $C_{1}^{PP}(x,y)$ is suppressed w.r.t.~$C_{2}^{PP}(x,y)$, since only the latter function contains the ``large'' integral $I_{M}(x-y)$. This is consistent with the representation in the chiral limit, given in Eq.~(\ref{eq:TPPpipi}).\\
For (iso-)scalar operators, we find at leading order accuracy that $C^{00}_{PP}=0$, $C_{1,2,3}^{SS}=0\,$ and
\begin{equation}\label{eq:resC00SS}
C^{00}_{SS} = 8B^{2}e^{i\frac{\Delta}{2}(x+y)}\cos\left(\bar{p}\cdot(x-y)\right)I_{\pi}(x-y)\,.
\end{equation}
The pion-exchange contributions to vector-isovector matrix elements are given by
\begin{eqnarray}
(C_{1}^{VV})^{\mu\nu} &=& 2e^{i\frac{\Delta}{2}(x+y)}\cos\left(\bar{p}\cdot(x-y)\right)\biggl(g^{\mu\nu}I_{\pi}^{(2)} + (x-y)^{\mu}(x-y)^{\nu}I_{\pi}^{(3)} \nonumber \\ &+& \left(\Delta^{\mu}(x-y)^{\nu}-\Delta^{\nu}(x-y)^{\mu}\right)I_{\pi}^{(1)} + \left(\bar{p}^{\mu}\bar{p}^{\nu}-\frac{1}{4}\Delta^{\mu}\Delta^{\nu}\right)I_{\pi}\biggr) \label{eq:resC1VV} \\ 
 &+& e^{i\frac{\Delta}{2}(x+y)}i\sin\left(\bar{p}\cdot(x-y)\right)\biggl(\left(\bar{p}^{\nu}\Delta^{\mu}-\bar{p}^{\mu}\Delta^{\nu}\right)I_{\pi} + 2\left(\bar{p}^{\mu}(x-y)^{\nu}+\bar{p}^{\nu}(x-y)^{\mu}\right)I_{\pi}^{(1)}\biggr)\,,\nonumber \\ 
(C_{2}^{VV})^{\mu\nu} &=& -\frac{3}{2}(C_{1}^{VV})^{\mu\nu}\,,\label{eq:resC2VV}\\
(C_{3}^{VV})^{\mu\nu} &=& -e^{i\frac{\Delta}{2}(x+y)}i\sin\left(\bar{p}\cdot(x-y)\right)\biggl(g^{\mu\nu}I_{\pi}^{(2)} + (x-y)^{\mu}(x-y)^{\nu}I_{\pi}^{(3)} \nonumber \\ &+& \left(\Delta^{\mu}(x-y)^{\nu}-\Delta^{\nu}(x-y)^{\mu}\right)I_{\pi}^{(1)} + \left(\bar{p}^{\mu}\bar{p}^{\nu}-\frac{1}{4}\Delta^{\mu}\Delta^{\nu}\right)I_{\pi}\biggr) \label{eq:resC3VV} \\ 
 &-& e^{i\frac{\Delta}{2}(x+y)}\cos\left(\bar{p}\cdot(x-y)\right)\biggl(\frac{1}{2}\left(\bar{p}^{\nu}\Delta^{\mu}-\bar{p}^{\mu}\Delta^{\nu}\right)I_{\pi} + \left(\bar{p}^{\mu}(x-y)^{\nu}+\bar{p}^{\nu}(x-y)^{\mu}\right)I_{\pi}^{(1)}\biggr)\,.\nonumber
\end{eqnarray}
$(C_{3}^{VV})^{\mu\nu}$ vanishes in the limit $p,p'\rightarrow 0$, so that the isospin structure of the matrix element is compatible with the low-energy theorem in Eq.~(\ref{eq:TVVpipi}).

\newpage

\newpage

For the axial-vector case, we find
\begin{eqnarray}
(C_{1}^{AA})^{\mu\nu} &=& -e^{i\frac{\Delta}{2}(x+y)}\biggl(\cos\left(\frac{\Delta}{2}\cdot(x-y)\right)\left(2I^{\mu\nu}_{\pi} -\left((x-y)^{\mu}\Delta^{\nu}-(x-y)^{\nu}\Delta^{\mu}\right)I^{(1)}_{\pi}\right) \nonumber \\ &+& i\sin\left(\frac{\Delta}{2}\cdot(x-y)\right)\left((x-y)^{\mu}\Delta^{\nu}+(x-y)^{\nu}\Delta^{\mu}\right)I^{(1)}_{\pi}\biggr) \\
 &+& e^{i\frac{\Delta}{2}(x+y)}\left(M_{\pi}^{2}-\Delta^2\right)\biggl(I^{\mu\nu}_{\pi\pi}-\frac{1}{4}\Delta^{\mu}\Delta^{\nu}I_{\pi\pi}^{\Delta} + \frac{1}{2}\left((x-y)^{\mu}\Delta^{\nu}-(x-y)^{\nu}\Delta^{\mu}\right)I^{(1)}_{\pi}\biggr)\,,\nonumber \\
(C_{2}^{AA})^{\mu\nu} &=& e^{i\frac{\Delta}{2}(x+y)}\cos\left(\frac{\Delta}{2}\cdot(x-y)\right)\left(3I^{\mu\nu}_{\pi} -\frac{3}{2}\left((x-y)^{\mu}\Delta^{\nu}-(x-y)^{\nu}\Delta^{\mu}\right)I^{(1)}_{\pi}\right)  \nonumber \\
 &+&  e^{i\frac{\Delta}{2}(x+y)}\biggl(\frac{3}{2}i\sin\left(\frac{\Delta}{2}\cdot(x-y)\right)\left((x-y)^{\mu}\Delta^{\nu}+(x-y)^{\nu}\Delta^{\mu}\right)I^{(1)}_{\pi}\nonumber \\
 &-& \left(2M_{\pi}^{2}-\frac{3}{2}\Delta^2\right)\left(I^{\mu\nu}_{\pi\pi}-\frac{1}{4}\Delta^{\mu}\Delta^{\nu}I^{\Delta}_{\pi\pi} + \frac{1}{2}\left((x-y)^{\mu}\Delta^{\nu}-(x-y)^{\nu}\Delta^{\mu}\right)I^{(1)}_{\pi}\right) \biggr)\,,\\
(C_{3}^{AA})^{\mu\nu} &=& 2e^{i\frac{\Delta}{2}(x+y)}\biggl(g^{\mu\nu}\bar{p}\cdot(x-y)I^{\Delta(7)}_{\pi\pi} + (x-y)^{\mu}(x-y)^{\nu}\bar{p}\cdot(x-y)I^{\Delta(8)}_{\pi\pi} \nonumber \\
 &+& \bar{p}^{\mu}(x-y)^{\nu}\left(I^{\Delta(7)}_{\pi\pi} - e^{i\frac{\Delta}{2}\cdot(x-y)}I^{(1)}_{\pi}\right) + \bar{p}^{\nu}(x-y)^{\mu}\left(I^{\Delta(7)}_{\pi\pi} - e^{-i\frac{\Delta}{2}\cdot(x-y)}I^{(1)}_{\pi}\right) \nonumber \\ &+& \bar{p}^{\mu}\Delta^{\nu}\left(I^{\Delta(9)}_{\pi\pi}+\frac{1}{2}I^{\Delta(3)}_{\pi\pi}\right) + \bar{p}^{\nu}\Delta^{\mu}\left(I^{\Delta(9)}_{\pi\pi}-\frac{1}{2}I^{\Delta(3)}_{\pi\pi}\right) \nonumber \\
 &+& (x-y)^{\mu}\Delta^{\nu}\bar{p}\cdot(x-y)\left(I^{\Delta(11)}_{\pi\pi}+\frac{1}{2}I^{\Delta(4)}_{\pi\pi}\right) + (x-y)^{\nu}\Delta^{\mu}\bar{p}\cdot(x-y)\left(I^{\Delta(11)}_{\pi\pi}-\frac{1}{2}I^{\Delta(4)}_{\pi\pi}\right) \nonumber \\ &+& \Delta^{\mu}\Delta^{\nu}\bar{p}\cdot(x-y)\left(I^{\Delta(12)}_{\pi\pi}-\frac{1}{4}I^{\Delta(1)}_{\pi\pi}\right)\biggr)\,.
\end{eqnarray}
In the expressions for the $VV$ and $AA$ correlators, we have left out the argument $x-y$ of the Fourier integrals for brevity. While in the latter two cases, the predictions are parameter-free, the PP and SS correlators depend on the QCD renormalization scale via the low-energy parameter $B$. Numerical estimates for this parameter at $\mu_{QCD}=2\,\mathrm{GeV}$ can be found e.~g. in the review \cite{Colangelo:2010et}.

\subsection{Results for $\vec{p}=\vec{p'}=0$}

In the case of zero four-momentum transfer, $p^{\mu}=p'^{\mu}=\bar{p}^{\mu},\,\Delta^{\mu}=0$, with $\bar{p}^{\mu}=(M_{\pi},\vec{0})$, and $r:= \sqrt{-(x-y)^2}$, the above results simplify considerably:
\begin{eqnarray}
C_{1}^{PP} &=& -\frac{B^{2}M_{\pi}^{2}}{2\pi^2}K_{0}(M_{\pi}r)\,,\quad C_{2}^{PP} = -\frac{B^{2}M_{\pi}^{2}}{\pi^2}\left(\frac{K_{1}(M_{\pi}r)}{M_{\pi}r} -K_{0}(M_{\pi}r)\right)\,,\label{eq:resC1PPDelta0}\\
C_{3}^{PP} &=& -i\frac{B^{2}M_{\pi}^{2}}{\pi^2}\frac{(x^0-y^0)}{r}K_{1}(M_{\pi}r)\,,\label{eq:resC3PPDelta0}\\
C^{00}_{SS} &=& \frac{2B^{2}M_{\pi}}{\pi^2r}\cos\left(M_{\pi}(x^0-y^0)\right)K_{1}(M_{\pi}r)\,,\label{eq:resC00SSDelta0}\\
\end{eqnarray}
\begin{eqnarray}
(C_{1}^{VV})^{\mu\nu} &=& -\frac{M_{\pi}^{2}\cos\left(M_{\pi}(x^0-y^0)\right)}{2\pi^2r^2}\biggl(g^{\mu\nu}K_{2}(M_{\pi}r) + (x-y)^{\mu}(x-y)^{\nu}\frac{M_{\pi}}{r}K_{3}(M_{\pi}r) \nonumber \\ 
 &-& \bar{p}^{\mu}\bar{p}^{\nu}\frac{r}{M_{\pi}}K_{1}(M_{\pi}r)\biggr)\label{eq:resC1VVDelta0} \\  
 &-& \frac{M_{\pi}^{2}\sin\left(M_{\pi}(x^0-y^0)\right)}{2\pi^2r^2}\left(\bar{p}^{\mu}(x-y)^{\nu}+\bar{p}^{\nu}(x-y)^{\mu}\right)K_{2}(M_{\pi}r)\,,\nonumber \\ 
(C_{2}^{VV})^{\mu\nu} &=& -\frac{3}{2}(C_{1}^{VV})^{\mu\nu}\,,\label{eq:resC2VVDelta0}\\
(C_{3}^{VV})^{\mu\nu} &=& i\frac{M_{\pi}^{2}\sin\left(M_{\pi}(x^0-y^0)\right)}{4\pi^2r^2}\biggl(g^{\mu\nu}K_{2}(M_{\pi}r) + (x-y)^{\mu}(x-y)^{\nu}\frac{M_{\pi}}{r}K_{3}(M_{\pi}r) \nonumber \\ 
 &-& \bar{p}^{\mu}\bar{p}^{\nu}\frac{r}{M_{\pi}}K_{1}(M_{\pi}r)\biggr) \label{eq:resC3VVDelta0} \\ 
 &-& i\frac{M_{\pi}^{2}\cos\left(M_{\pi}(x^0-y^0)\right)}{4\pi^2r^2}\left(\bar{p}^{\mu}(x-y)^{\nu}+\bar{p}^{\nu}(x-y)^{\mu}\right)K_{2}(M_{\pi}r)\,,\nonumber\\
(C_{1}^{AA})^{\mu\nu} &=& g^{\mu\nu}\frac{M_{\pi}^{3}}{8\pi^2r}\left(K_{1}(M_{\pi}r) + \frac{4}{M_{\pi}r}K_{2}(M_{\pi}r)\right) \nonumber \\ 
 &+& (x-y)^{\mu}(x-y)^{\nu}\frac{M_{\pi}^{4}}{8\pi^2r^2}\left(K_{2}(M_{\pi}r)+\frac{4}{M_{\pi}r}K_{3}(M_{\pi}r)\right)\,,\label{eq:resC1AADelta0}\\
(C_{2}^{AA})^{\mu\nu} &=& -g^{\mu\nu}\frac{M_{\pi}^{3}}{4\pi^2r}\left(K_{1}(M_{\pi}r) + \frac{3}{M_{\pi}r}K_{2}(M_{\pi}r)\right) \nonumber \\
 &-& (x-y)^{\mu}(x-y)^{\nu}\frac{M_{\pi}^{4}}{4\pi^2r^2}\left(K_{2}(M_{\pi}r)+\frac{3}{M_{\pi}r}K_{3}(M_{\pi}r)\right)\,,\label{eq:resC2AADelta0}\\
(C_{3}^{AA})^{\mu\nu} &=& i\left(M_{\pi}(x^0-y^0)g^{\mu\nu}-\bar{p}^{\mu}(x-y)^{\nu}-\bar{p}^{\nu}(x-y)^{\mu}\right)\frac{M_{\pi}^{2}}{4\pi^2r^2}K_{2}(M_{\pi}r) \nonumber \\
 &+& i(x^0-y^0)(x-y)^{\mu}(x-y)^{\nu}\frac{M_{\pi}^{4}}{4\pi^2r^3}K_{3}(M_{\pi}r)\,.\label{eq:resC3AADelta0} 
\end{eqnarray}
Here, the $K_{i}(z)$ are the modified Bessel functions of the second kind, see also App.~\ref{app:integrale}.
For the previous expressions it is straightforward to verify that
\begin{eqnarray}
\partial_{\mu}^{x}(C_{i}^{VV})^{\mu\nu} &=& 0 = \partial_{\nu}^{y}(C_{i}^{VV})^{\mu\nu}\,,\label{eq:cVconserved}\\
\partial_{\mu}^{x}\partial_{\nu}^{y}(C_{i}^{AA})^{\mu\nu} &=& \frac{1}{4}\left(\frac{M_{\pi}^{2}}{B}\right)^{2}C_{i}^{PP}\,,\quad i=1,2,3\,.\label{eq:PCAC}
\end{eqnarray}
The first equation states the conservation of the vector current in the case of perfect isospin symmetry, while the second equation is a direct consequence of the well-known PCAC relation for the axial currents. We also note that, in the chiral limit $M_{\pi}\rightarrow 0$, the zero-momentum results reduce to
\begin{equation}\label{eq:chlim1}
C_{1,3}^{PP}\rightarrow 0\,,\quad C_{2}^{PP}\rightarrow -\left(\frac{B}{\pi r}\right)^{2}\leftarrow -\frac{1}{2}C_{SS}^{00}\,,
\end{equation}
in accord with the general theorems in Eqs.~(\ref{eq:TS0S0pipi}), (\ref{eq:TPPpipi}), and 
\begin{eqnarray}
(C_{2}^{VV})^{\mu\nu} &=& -\frac{3}{2}(C_{1}^{VV})^{\mu\nu}\rightarrow \frac{3}{2\pi^2r^4}\left(g^{\mu\nu}+\frac{4}{r^2}(x-y)^{\mu}(x-y)^{\nu}\right) \leftarrow -(C_{2}^{AA})^{\mu\nu}\leftarrow \frac{3}{2}(C_{1}^{AA})^{\mu\nu}\,,\nonumber \\
(C_{3}^{VV})^{\mu\nu} &\rightarrow& 0 \leftarrow -(C_{3}^{AA})^{\mu\nu}\,,\label{eq:chlim2}
\end{eqnarray}
in accord with Eqs.~(\ref{eq:TAApipi})-(\ref{eq:VV+AA=0}). This limiting value is of little practical use, however, because we can expect ChPT to give reliable predictions only for $M_{\pi}r\gtrsim 1$.

\newpage

\section{Resonance exchange}

The pion exchange contributions calculated in the previous section are expected to dominate at large distances $r\gg M_{\pi}^{-1}$. For practical applications of those expressions, 
it is of interest to quantify the relevance of higher-order contributions in this range of space-like distances. Instead of performing a full one-loop calculation in pion ChPT, we attempt here an estimate based on the exchange of the lowest-lying meson resonances relevant for the matrix elements considered in this work. Such a strategy has often been successful in hadron physics \cite{Sakurai:1960ju,Sakurai:1969ju,Weinberg:1967kj,Feynman:1973xc,Meissner:1987ge,Ecker:1988te}, and has lead (in connection with ChPT and the large-$N_{c}$ limit of QCD \cite{'tHooft:1973jz}) to the so-called ``resonance chiral theory'' \cite{Bernard:1991zc,Shabalin:1998im,RuizFemenia:2003gw,Cirigliano:2004ue,Rosell:2004mn,Rosell:2006dt,SanzCillero:2009ap,SanzCillero:2010gy,Guo:2014yva}.

\subsection{Chiral Lagrangians for resonances}

Let us start with spin 1 meson resonances, of which the $\rho$ is probably the most relevant example.
To describe the exchange of vector mesons between $x$ und $y$, we employ the vector field formalism of \cite{Ecker:1989yg} (the vector field $V^{\mu}$ contains the $\rho$ fields). We remark that, in the antisymmetric tensorfield formalism \cite{Ecker:1988te}, an additional contact term is generated, which is however only relevant for $x-y\rightarrow 0$, while here we are only interested in the behavior at large distances. The Lagrangian density for the interaction of vector mesons with pions and external vector and axial-vector source fields is (to lowest order) given by \cite{Ecker:1989yg}
\begin{equation}\label{eq:LV}
\mathcal{L}_{V}^{\mathrm{int}} = -\frac{f_{V}}{2\sqrt{2}}\langle F^{+}_{\mu\nu}V^{\mu\nu}\rangle - \frac{ig_{V}}{2\sqrt{2}}\langle\lbrack u_{\mu},\,u_{\nu}\rbrack V^{\mu\nu}\rangle + \ldots\,,
\end{equation}
\begin{eqnarray}
V^{\mu\nu} &=& D^{\mu}V^{\nu}-D^{\nu}V^{\mu}:=\partial^{\mu}V^{\nu}-\partial^{\nu}V^{\mu} + \lbrack\Gamma^{\mu},\,V^{\nu}\rbrack - \lbrack\Gamma^{\nu},\,V^{\mu}\rbrack\,,\nonumber \\
\Gamma^{\mu} &=& \frac{1}{2}\left(u^{\dagger}[\partial^{\mu}-i(v^{\mu}+a^{\mu})]u + u[\partial^{\mu}-i(v^{\mu}-a^{\mu})]u^{\dagger}\right)\,,\nonumber \\
F^{\pm}_{\mu\nu} &=& uF^{L}_{\mu\nu}u^{\dagger} \pm u^{\dagger}F^{R}_{\mu\nu}u\,,\quad u_{\mu}=iu^{\dagger}\left(\nabla_{\mu}U\right)u^{\dagger}\,,\label{eq:resbuildingblocks} \\
F^{R,L}_{\mu\nu} &=& \partial_{\mu}(v_{\nu}\pm a_{\nu})-\partial_{\nu}(v_{\mu}\pm a_{\mu})-i\lbrack(v_{\mu}\pm a_{\mu}),\,(v_{\nu}\pm a_{\nu})\rbrack\,,\nonumber \\
\nabla_{\mu}U &=& \partial_{\mu}U-i(v_{\mu}+a_{\mu})U+iU(v_{\mu}-a_{\mu})\,,\quad u=\sqrt{U}\,,\nonumber \\
V_{\mu} &=& \rho^{0}_{\mu}\bar{\lambda}^{\rho^{0}} + \rho^{+}_{\mu}\bar{\lambda}^{\rho^{+}} + \rho^{-}_{\mu}\bar{\lambda}^{\rho^{-}}\,,\quad \bar{\lambda}^{\rho^{0}}=\bar{\lambda}^{\pi^{0}}\,,\quad \bar{\lambda}^{\rho^{\pm}}=\bar{\lambda}^{\pi^{\pm}}\,,\nonumber 
\end{eqnarray}
see Eq.~(\ref{eq:channelmpi}) for the channel matrices $\bar{\lambda}$. The vector field propagator in momentum space is
\begin{equation}
D_{\alpha\beta}(q) = (-i)\frac{g_{\alpha\beta}-\frac{q_{\alpha}q_{\beta}}{M_{V}^{2}}}{q^2-M_{V}^{2}}\,.
\end{equation}
The chiral Lagrangian for axial-vector resonances is at leading order (see.~e.~g.~\cite{Ecker:1989yg}, Eq.~(52))
\begin{eqnarray}
\mathcal{L}_{A}^{\mathrm{int}} &=&  -\frac{f_{A}}{2\sqrt{2}}\langle F^{-}_{\mu\nu}A^{\mu\nu}\rangle + \ldots\,,\label{eq:LA}\\
A^{\mu\nu} &=& D^{\mu}\xi^{\nu}-D^{\nu}\xi^{\mu}:=\partial^{\mu}\xi^{\nu}-\partial^{\nu}\xi^{\mu} + \lbrack\Gamma^{\mu},\,\xi^{\nu}\rbrack - \lbrack\Gamma^{\nu},\,\xi^{\mu}\rbrack\,.
\end{eqnarray}
The $\xi^{\mu}$ contain the isovector axial-vector fields (called $a_{1}$). We show only the terms relevant for the contributions we calculate below - for a complete list of terms, see Eqs.~(15), (49) in \cite{Ecker:1989yg}.\\
In addition we consider chiral couplings of $I=1$ scalar and $I=0$ pseudoscalar resonances (denoted here as $a_{0}$ and $\eta$, respectively) and an isosinglet scalar field called $\sigma$ (compare e.~g. \cite{Pelaez:2015qba} for an up-to-date review on the lowest-lying scalar resonance).
We do not include isovector pseudoscalar resonance fields, since the corresponding resonances have masses $\gtrsim 1.3\,\mathrm{GeV}$ and are therefore expected to be less relevant for our purposes.
At lowest chiral order the pertaining Lagrangians are of the form (compare \cite{Ecker:1988te}, Eqs.~(3.14-15))
\begin{eqnarray}
\mathcal{L}_{S} &=& \frac{1}{2}\langle D_{\mu}a_{0}D^{\mu}a_{0} - M_{S}^{2}a_{0}^{2}\rangle + c_{d}\langle a_{0}u_{\mu}u^{\mu}\rangle + c_{m}\langle a_{0}\chi_{+}\rangle\,,\label{eq:LS}\\
\mathcal{L}_{P} &=& \frac{1}{2}\partial_{\mu}\eta \partial^{\mu}\eta - \frac{1}{2}M_{P}^{2}\eta^{2} + id_{\eta}\eta\langle\chi_{-}\rangle\,,\label{eq:LP}\\
\mathcal{L}_{\sigma} &=& \frac{1}{2}\partial_{\mu}\sigma \partial^{\mu}\sigma - \frac{1}{2}m_{\sigma}^{2}\sigma^{2} + c_{d}^{\sigma}\sigma\langle u_{\mu}u^{\mu}\rangle + c_{m}^{\sigma}\sigma\langle\chi_{+}\rangle\,,\label{eq:Lsigma}\\
D_{\mu}a_{0} &=& \partial_{\mu}a_{0} + \lbrack\Gamma_{\mu},\,a_{0}\rbrack\,,\quad u_{\mu}=iu^{\dagger}\left(\nabla_{\mu}U\right)u^{\dagger}\,,\quad \chi_{\pm} = 2B\left(u^{\dagger}(s+ip)u^{\dagger}\pm u(s-ip)u\right)\,,\nonumber \\
\Gamma_{\mu} &=& \frac{1}{2}\left(u^{\dagger}[\partial_{\mu}-i(v_{\mu}+a_{\mu})]u + u[\partial_{\mu}-i(v_{\mu}-a_{\mu})]u^{\dagger}\right)\,.\nonumber
\end{eqnarray}
It should be remarked that, although we borrow the notation from \cite{Ecker:1988te}, our Lagrangians above represent the two-flavor (chiral SU(2)) versions of the Lagrangians given in this reference.
Estimates for the parameters entering the resonance Lagrangians can be found in Eq.~(67) of \cite{Ecker:1989yg}, and Sec.~4 of \cite{Ecker:1988te}. For vector meson masses and couplings at higher quark masses, see also \cite{Chen:2015tpa} ($f_{\rho}=\sqrt{2}M_{V}f_{V}$). As the quark mass dependence is formally of higher order in the chiral expansion, we can e.~g. set $F\approx F_{\pi}$ (etc.). To produce rough estimates one can adopt
\begin{eqnarray}
F\approx F_{\pi}&\approx& 100\,\mathrm{MeV}\,,\quad F_{V}:=M_{V}f_{V}\approx 155\,\mathrm{MeV}\,,\nonumber\\
M_{V} \approx M_{\rho} &\approx& 800\,\mathrm{MeV}\,,\quad F_{A}:=M_{A}f_{A}\approx\frac{F_{V}}{\sqrt{2}}\,,\quad 2g_{V}\approx f_{V}\,.\label{eq:respars1}
\end{eqnarray}
The couplings entering the Lagrangians for (pseudo\,-)scalar resonances are not known to a satisfying accuracy, while the mass parameters can be approximated by the corresponding masses of the lowest-lying resonances. We take here $M_{P}\approx 600\,\,\mathrm{MeV}$. By a comparison with the chiral SU(3) calculation, we estimate the range
$|d_{\eta}|=\frac{F}{4\sqrt{3}}\approx 10\ldots 15\,\mathrm{MeV}$. $c_{m}$ should be (roughly) of order $\sim 50\,\mathrm{MeV}$, $c_{d}\sim 30\,\mathrm{MeV}$ (see also Eq.~(4) in \cite{Rosell:2006dt}, and Eqs.~(54,55) in \cite{Guo:2014yva}), $M_{S}\approx 1000\,\mathrm{MeV}$. From the sigma mass and width, and the comparison to the three-flavor calculation, we conclude that $c_{m}^{\sigma},c_{d}^{\sigma}$ should be of the same order of magnitude, $c_{m,d}^{\sigma}\sim c_{m,d}/\sqrt{2}$.\\
The reader might object that the use of the resonance Lagrangians specified above is not justified, because the Fourier integrals of the pole diagrams receive the dominant contribution from the region where the four-momenta $q,q'$ obey $q^2\sim M_{R}^{2}$ (where $R$ stands for the resonances), which is beyond the regime where the chiral power counting applies. While this is, strictly speaking, true, and the resonance-exchange terms are not the only higher-order corrections to the pion-exchange tree graphs, we note that the vertex structures obtained from the above Lagrangians are basically unique up to quark mass corrections, corrections due to the (soft) momenta $p,p'$ of the pions, and corrections of $\mathcal{O}(q^2-M_{R}^{2})$. For example, for $p,p'\rightarrow 0$, $M_{\pi}\rightarrow 0$, $q^2\sim q'^2\sim M_{R}^{2}$, the coupling of $s^{0}$ to the $\sigma$ is just a constant, while the vertex structures for the couplings of $v^{a},a^{a}$ to the (axial-)vector resonances must be of the form $\sim g^{\mu\nu}q^{2}-q^{\mu}q^{\nu}$ (up to field transformations \cite{Chisholm:1961tha}) to ensure that the correct number of degrees of freedom for a spin-1 particle propagates (transversality of the coupling, see e.~g.~(\ref{eq:vrhovertex})). For a given term in the Lagrangian, the relative strength of the couplings is fixed (by construction) by chiral symmetry. One should also be aware of the fact that the coupling constants $f_{V},\,g_{V}$ etc. are also determined from processes where the resonance four-momenta are close to $q^2\sim M_{R}^{2}$, and not soft in the sense of ChPT. Strictly speaking, the chiral power counting of Ref.~\cite{Ecker:1988te} applies only when the resonance is strongly virtual, $|q^2|\ll M_{R}^{2}$. The problem becomes even more severe when loop graphs with resonance propagators are considered. For a discussion of such issues, we refer to \cite{Rosell:2004mn,Bruns:2004tj,Bruns:2013tja,Fuchs:2003sh,Djukanovic:2009zn}, and references therein.

\subsection{Resonance exchange contributions}

\begin{figure}[h]
\centering
\includegraphics[width=0.49\textwidth]{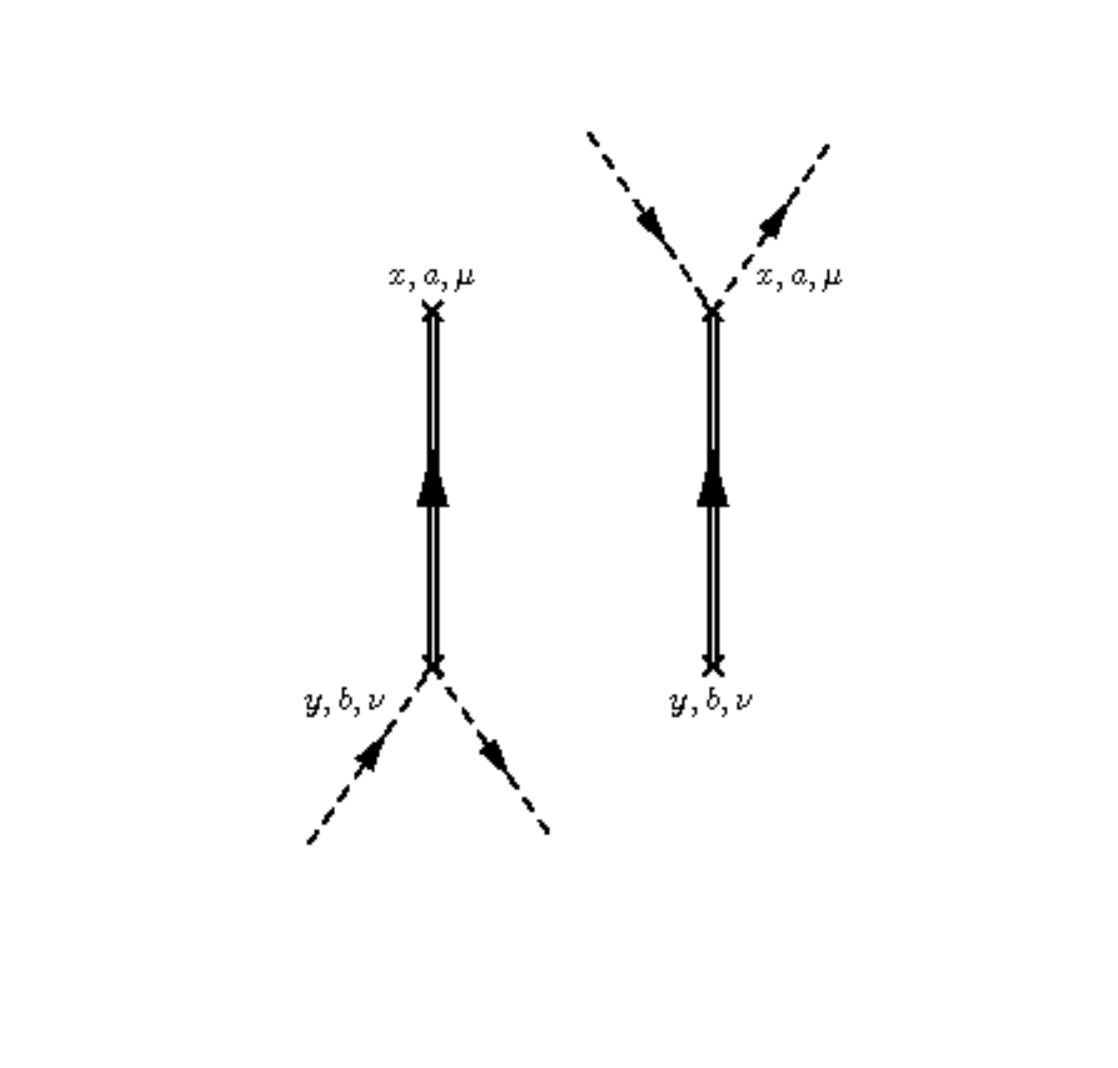}
\caption{Resonance (double line) exchange graphs of ``T''-topology.}
\label{fig:fmgraphsRhoVV}
\end{figure}

\begin{figure}[h]
\centering
\includegraphics[width=0.49\textwidth]{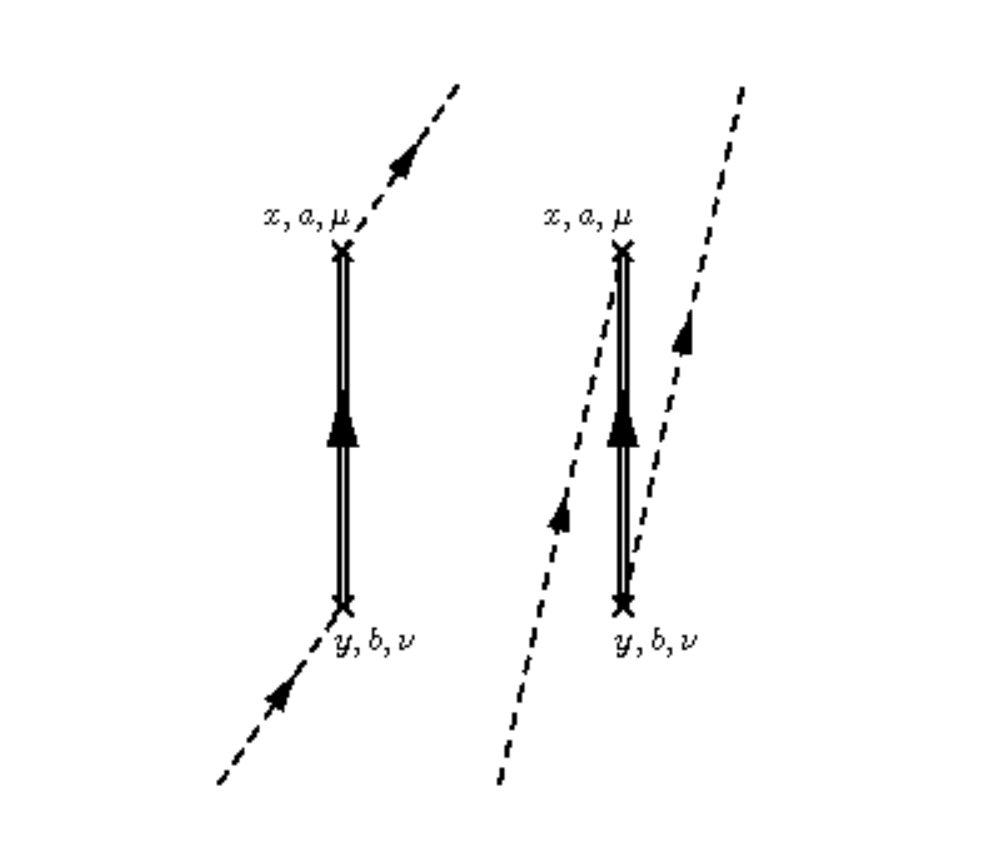}
\caption{Resonance (double line) exchange graphs of ``Z''-topology.}
\label{fig:fmgraphsRhoAA}
\end{figure}

As explained above, we evaluate some specific higher-order conributions which have the form of resonance pole-diagrams, to investigate how the leading $r-$dependence, given by the tree-level ChPT-contributions calculated in (\ref{eq:resC1PP})-(\ref{eq:resC3AADelta0}), is possibly modified, and for which range of distances $|\vec{x}-\vec{y}|$ those formulae yield reliable predictions. To this end, it is sufficient to evaluate the resonance exchange contributions for vanishing pion momenta, and in the chiral limit $M_{\pi}\rightarrow 0$ (this limit is not critical here, because those graphs come without pion propagators). The corresponding graph topologies are depicted in Figs.~\ref{fig:fmgraphsRhoVV},\ref{fig:fmgraphsRhoAA}. Contributions of X-topology, which contain a $\pi$-resonance scattering vertex, are chirally suppressed and do not contribute in the limit considered here.\\
The $\sigma$ exchange graphs contribute as
\begin{equation}
C_{SS(R)}^{00}(x-y) = -\frac{128B^{2}(c_{m}^{\sigma})^{2}}{F^{2}}I_{\sigma}(x-y)\,,\quad C_{2(R)}^{PP}(x-y) = \frac{64B^{2}(c_{m}^{\sigma})^{2}}{F^{2}}I_{\sigma}(x-y)\,,
\end{equation}
while the contributions to $C_{1,3}^{PP}$ vanish (as they should in the chiral limit, see (\ref{eq:TPPpipi})). Here the label $(R)$ marks the resonance contribution to the correlator, and $I_{\sigma}(x-y)$ is given by the integral $I_{M}(x-y)$ defined in (\ref{eq:defIM}), with $M\rightarrow m_{\sigma}$ (and similarly for the other resonances). 
The $\eta$ and $a_{0}$ exchange graphs lead to
\begin{eqnarray}
C_{2(R)}^{SS}(x-y) &=& \frac{32B^{2}}{F^{2}}\left(2d_{\eta}^{2}I_{\eta}(x-y)-c_{m}^{2}I_{a_{0}}(x-y)\right)\,,\quad C_{1,3(R)}^{SS}(x-y) = 0\,, \\
C_{PP(R)}^{00}(x-y) &=& -\frac{64B^{2}}{F^{2}}\left(2d_{\eta}^{2}I_{\eta}(x-y)-c_{m}^{2}I_{a_{0}}(x-y)\right)\,.
\end{eqnarray}
The $\rho$ and $a_{1}$ exchange graphs contribute to the isovector $VV$ and $AA$ correlators:
\begin{eqnarray}
(C_{1(R)}^{VV})^{\mu\nu}(x-y) &=& \frac{2(f_{V}M_{V})^{2}}{F^{2}}\left(g^{\mu\nu}\left(M_{V}^{2}I_{V}(x-y)-I_{V}^{(2)}(x-y)\right) - (x-y)^{\mu}(x-y)^{\nu}I_{V}^{(3)}(x-y)\right) \nonumber \\
 &-& \frac{2(f_{A}M_{A})^{2}}{F^{2}}\left(g^{\mu\nu}\left(M_{A}^{2}I_{A}(x-y)-I_{A}^{(2)}(x-y)\right) - (x-y)^{\mu}(x-y)^{\nu}I_{A}^{(3)}(x-y)\right)\,,\nonumber \\
(C_{2(R)}^{VV})^{\mu\nu}(x-y) &=& -\frac{3}{2}(C_{1(R)}^{VV})^{\mu\nu}(x-y)\,,\quad (C_{3(R)}^{VV})^{\mu\nu}(x-y) = 0\,.
\end{eqnarray}
The coefficient functions $I_{M}^{(2,3)}$ are defined in Eq.~(\ref{eq:IMtensordef}). For the axial-vector correlator, we find $(C_{i(R)}^{AA})^{\mu\nu}=-(C_{i(R)}^{VV})^{\mu\nu}(x-y)$, as expected in the chiral and zero momentum limit (see Eq.~(\ref{eq:VV+AA=0})).

\newpage

\newpage

\begin{figure}[!h]
\centering
\subfigure[\,$C_{SS}^{++}(M_{\pi}=300\,\mathrm{MeV})$]{\includegraphics[width=0.44\textwidth]{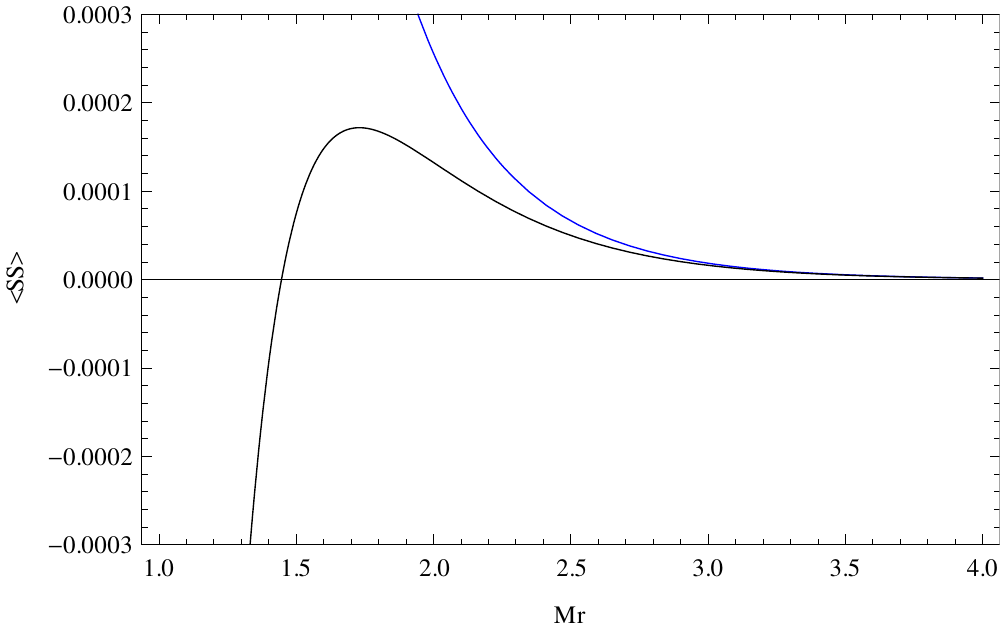}}
\subfigure[\,$C_{SS}^{++}(M_{\pi}=150\,\mathrm{MeV})$]{\includegraphics[width=0.44\textwidth]{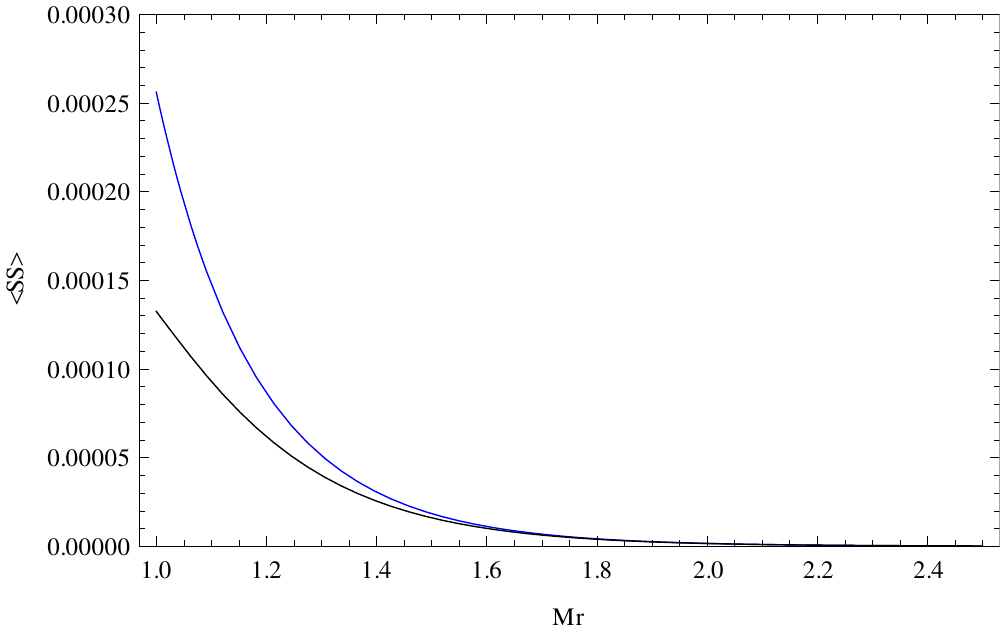}}\\
\subfigure[\,$C_{PP}^{++}(M_{\pi}=300\,\mathrm{MeV})$]{\includegraphics[width=0.44\textwidth]{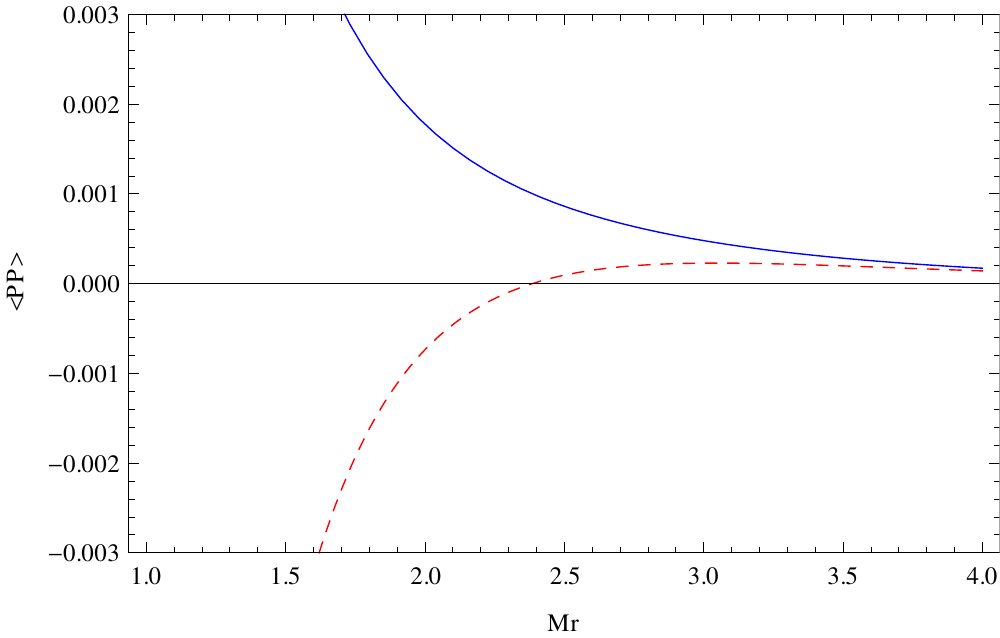}}
\subfigure[\,$C_{PP}^{++}(M_{\pi}=150\,\mathrm{MeV})$]{\includegraphics[width=0.44\textwidth]{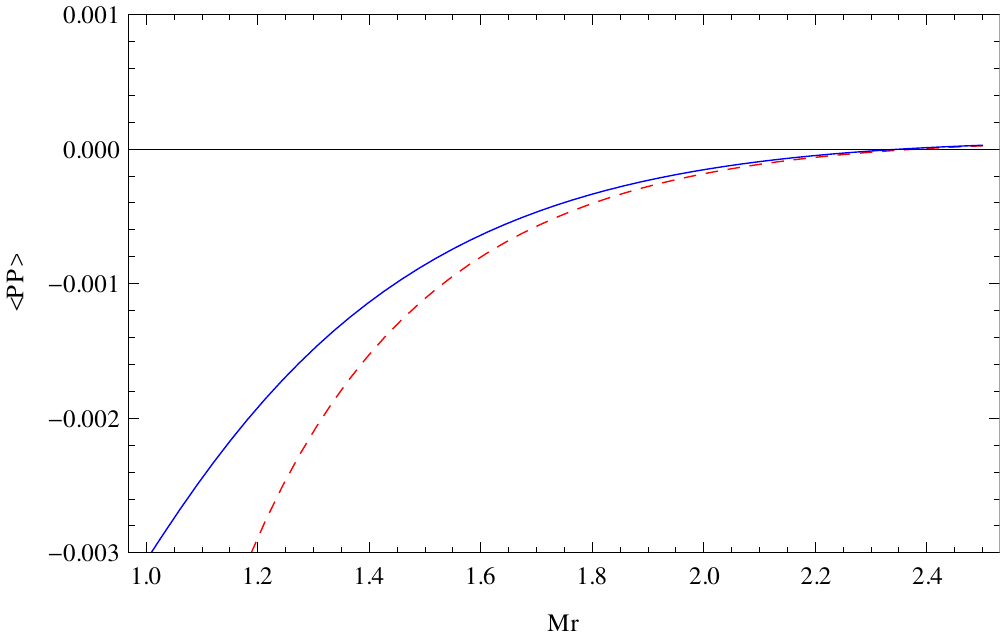}}
\caption{$C_{\mathcal{O}\mathcal{O}}^{++}$ for $M_{\pi}=300\,\mathrm{MeV}$ (left) and $M_{\pi}=150\,\mathrm{MeV}$ (right), in $\mathrm{GeV}^{4}$, as a function of $M_{\pi}r$. Red dashed lines: LO ChPT, blue: LO ChPT + first resonance, black: LO ChPT + first and second resonance.}
\label{fig:CPPSSwithR}
\end{figure}

\begin{figure}[!h]
\centering
\subfigure[\,$C_{VV}^{++}(M_{\pi}=300\,\mathrm{MeV})$]{\includegraphics[width=0.44\textwidth]{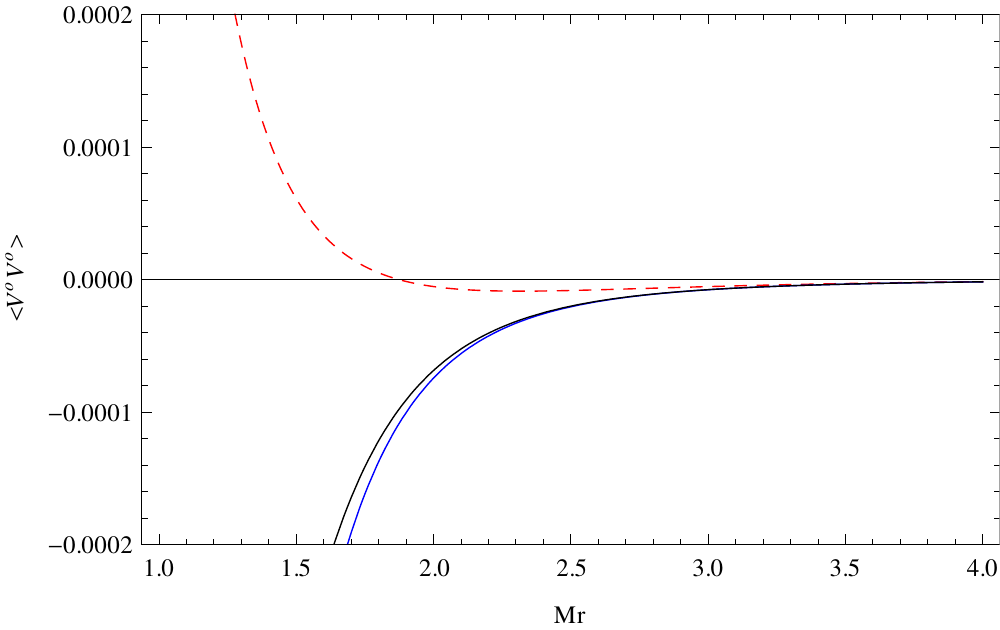}}
\subfigure[\,$C_{VV}^{++}(M_{\pi}=150\,\mathrm{MeV})$]{\includegraphics[width=0.44\textwidth]{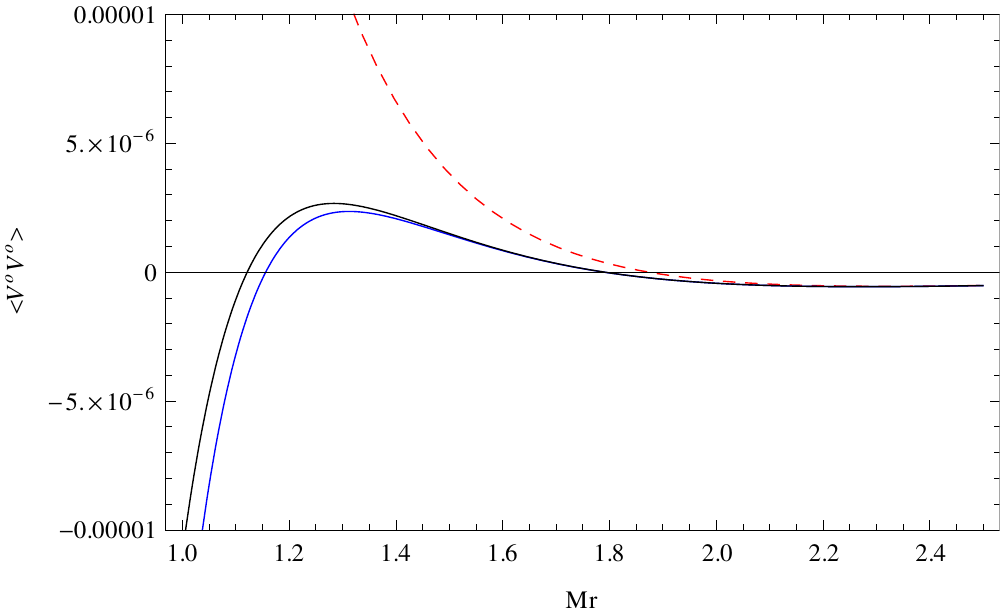}}\\
\subfigure[\,$C_{AA}^{++}(M_{\pi}=300\,\mathrm{MeV})$]{\includegraphics[width=0.44\textwidth]{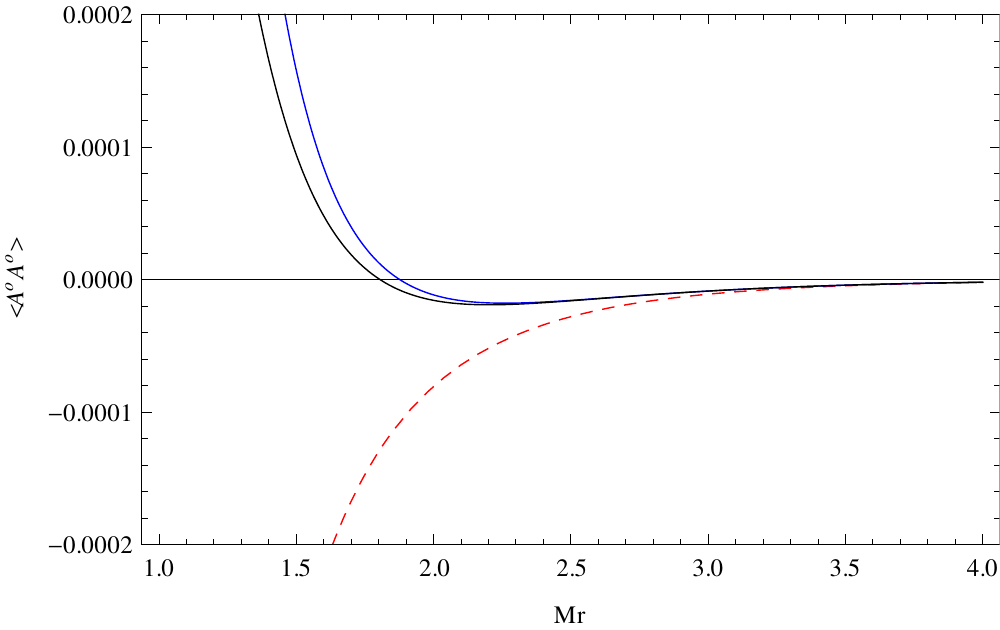}}
\subfigure[\,$C_{AA}^{++}(M_{\pi}=150\,\mathrm{MeV})$]{\includegraphics[width=0.44\textwidth]{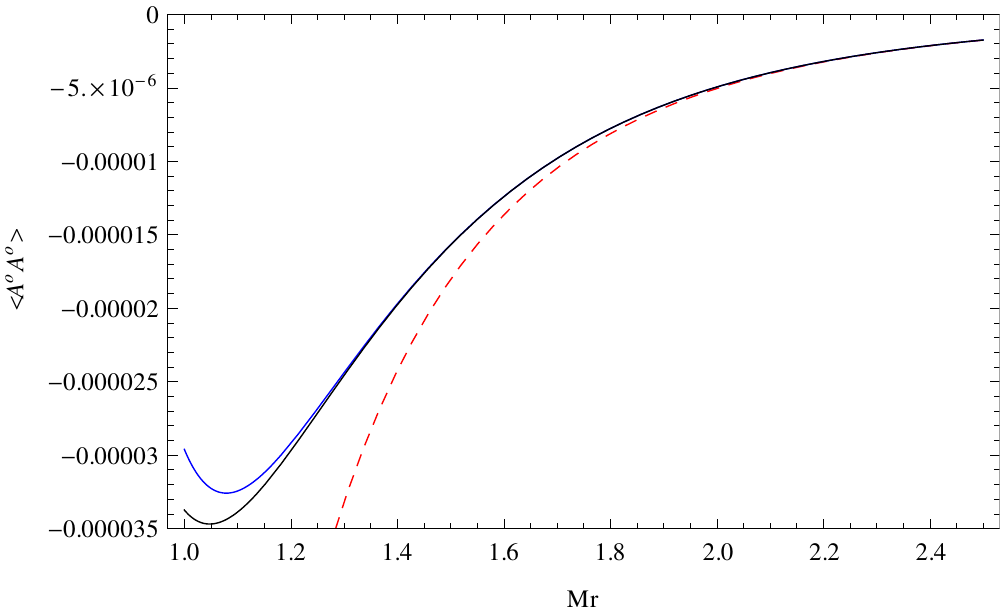}}
\caption{$C_{\mathcal{O}\mathcal{O}}^{++}$ for $M_{\pi}=300\,\mathrm{MeV}$ (left) and $M_{\pi}=150\,\mathrm{MeV}$ (right), in $\mathrm{GeV}^{4}$, as a function of $M_{\pi}r$. Red dashed lines: LO ChPT, blue: LO ChPT + first resonance, black: LO ChPT + first and second resonance.}
\label{fig:CVVAAwithR}
\end{figure}

As a specific example, let us consider the $\pi^{-}\rightarrow\pi^{+}$ matrix elements with two $\bar{u}(\ldots)d$ operators, for vanishing pion momenta, $x^{0}=y^{0}$ (so that $r=|\vec{x}-\vec{y}|$) and $\mu=\nu=0\,$:
\begin{eqnarray*}
C_{SS}^{++}(x-y) &:=& \langle\pi^{+}(p')|\bar{u}(x)d(x)\bar{u}(y)d(y)|\pi^{-}(p)\rangle = C_{1}^{SS}(x-y)+C_{2}^{SS}(x-y)\,, \\
C_{PP}^{++}(x-y) &:=& \langle\pi^{+}(p')|i\bar{u}(x)\gamma_{5}d(x)i\bar{u}(y)\gamma_{5}d(y)|\pi^{-}(p)\rangle = C_{1}^{PP}(x-y)+C_{2}^{PP}(x-y)\,, \\
C_{VV}^{++}(x-y) &:=& \langle\pi^{+}(p')|\bar{u}(x)\gamma_{0}d(x)\bar{u}(y)\gamma_{0}d(y)|\pi^{-}(p)\rangle  = 4(C_{1}^{VV})^{00}(x-y)+4(C_{2}^{VV})^{00}(x-y)\,, \\
C_{AA}^{++}(x-y) &:=& \langle\pi^{+}(p')|\bar{u}(x)\gamma_{0}\gamma_{5}d(x)\bar{u}(y)\gamma_{0}\gamma_{5}d(y)|\pi^{-}(p)\rangle  = 4(C_{1}^{AA})^{00}(x-y)+4(C_{2}^{AA})^{00}(x-y)\,,
\end{eqnarray*}
For easy reference, we give the explicit expressions below:
\begin{eqnarray}
C_{SS}^{++}(x-y) &=& 0 + \frac{8B^{2}}{\pi^2F^{2}r}\left(2M_{P}d_{\eta}^{2}K_{1}(M_{P}r)-M_{S}c_{m}^{2}K_{1}(M_{S}r)\right)\,,\\
C_{PP}^{++}(x-y) &=& \frac{B^{2}M_{\pi}}{2\pi^2r}\left(M_{\pi}rK_{0}(M_{\pi}r)-2K_{1}(M_{\pi}r)\right) + \frac{16B^{2}(c_{m}^{\sigma})^{2}m_{\sigma}}{\pi^2F^{2}r}K_{1}(m_{\sigma}r)\,,\\
C_{VV}^{++}(x-y) &=& \frac{M_{\pi}^{2}}{\pi^2r^2}\left(K_{2}(M_{\pi}r)-M_{\pi}rK_{1}(M_{\pi}r)\right) - \frac{f_{V}^{2}M_{V}^{4}}{\pi^2F^{2}r^2}\left(K_{2}(M_{V}r)+M_{V}rK_{1}(M_{V}r)\right) \nonumber \\
 &+& \frac{f_{A}^{2}M_{A}^{4}}{\pi^2F^{2}r^2}\left(K_{2}(M_{A}r)+M_{A}rK_{1}(M_{A}r)\right)\,,\\
C_{AA}^{++}(x-y) &=& -\frac{M_{\pi}^{2}}{2\pi^2r^2}\left(2K_{2}(M_{\pi}r)+M_{\pi}rK_{1}(M_{\pi}r)\right) + \frac{f_{V}^{2}M_{V}^{4}}{\pi^2F^{2}r^2}\left(K_{2}(M_{V}r)+M_{V}rK_{1}(M_{V}r)\right) \nonumber \\
 &-& \frac{f_{A}^{2}M_{A}^{4}}{\pi^2F^{2}r^2}\left(K_{2}(M_{A}r)+M_{A}rK_{1}(M_{A}r)\right)\,.
\end{eqnarray}
The zero in the first line reminds us of the fact that there is no pion-exchange contribution to $C_{SS}^{++}$ at leading order. In Figs.~\ref{fig:CPPSSwithR} and \ref{fig:CVVAAwithR}, we plot the results for the correlators as a function of $M_{\pi}r$, for $M_{\pi}=300\,\mathrm{MeV}$ and $150\,\mathrm{MeV}$.

The parameters used here are $B=2.5\,\mathrm{GeV},\,F=100\,\mathrm{MeV},\,m_{\sigma}=0.5\,\mathrm{GeV}$,\,$M_{S}=1\,\mathrm{GeV},\,M_{P}=0.6\,\mathrm{GeV},\,M_{V}=0.8\,\mathrm{GeV},\,M_{A}=1.25\,\mathrm{GeV}$,\,$d_{\eta}=15\,\mathrm{MeV},c_{m}=50\,\mathrm{MeV},c_{m}^{\sigma}=35\,\mathrm{MeV},f_{V}=0.2,\,f_{A}=0.1$\,. We observe that, for $M_{\pi}=300\,\mathrm{MeV}$, the resonance contributions become important already for $M_{\pi}r\lesssim 3$, pointing to the fact that the tree-level ChPT results are not reliable for smaller values of $M_{\pi}r$. For $M_{\pi}=150\,\mathrm{MeV}$, $M_{\pi}r\gtrsim 2$ seems to be sufficient (but note that lattice artefacts due to the finite volume, the finite lattice spacing etc. have not been taken into account so far (see also \cite{Burkardt:1994pw})). The distance between the operator insertions should therefore not be smaller than $\sim 2\,\mathrm{fm}$ if LO ChPT is to be trusted. The effects due to the resonances are seen to become very pronounced as soon as smaller distances are tested, and it seems futile to work out pion-loop corrections in the standard framework of ChPT to improve the description in this region. Note that, in ChPT, the resonance effects are usually encoded in contact terms of the chiral Lagrangians for the Goldstone bosons. However, the Fourier transformation of a momentum-dependent local term will just generate contact-terms $\sim\delta^{3}(\vec{x}-\vec{y})$ in position-space. While this shows that the intermediate region $1\lesssim M_{\pi}r\lesssim 2$ is not amenable to an analysis employing ChPT, it is nonetheless interesting to see whether the result with added resonance exchange contributions (the black curves in Figs.~\ref{fig:CPPSSwithR} and \ref{fig:CVVAAwithR}) yield reasonable estimates for the measured correlators on the lattice in this region. A first comparison with preliminary lattice data showed that this seems indeed to be the case. This will be further discussed elsewhere \cite{RQCD}.
\acknowledgments{I thank G.~Bali, M.~Diehl, A.~Sch\"afer and C.~Zimmermann for useful hints and discussions. This work was supported by the Deutsche Forschungsgemeinschaft SFB/Transregio 55.}

\newpage

\begin{appendix}

\section{Feynman rules}
\label{app:frules}
\def\theequation{\Alph{section}.\arabic{equation}}
\setcounter{equation}{0}

We collect the Goldstone boson fields in the matrix
\begin{displaymath}
\phi = \pi^{0}\bar{\lambda}^{\pi^{0}}+\pi^{+}\bar{\lambda}^{\pi^{+}}+\pi^{-}\bar{\lambda}^{\pi^{-}}\,.
\end{displaymath}
The ``channel matrices'' are given, in terms of Pauli matrices, by
\begin{equation}\label{eq:channelmpi}
\bar{\lambda}^{\pi^{+}} = \frac{1}{2}(\tau^{1}+i\tau^{2})\,,\quad \bar{\lambda}^{\pi^{-}} = \frac{1}{2}(\tau^{1}-i\tau^{2})\,,\quad \bar{\lambda}^{\pi^{0}} = \frac{1}{\sqrt{2}}\tau^{3}\,.
\end{equation}
They obey the algebraic identity
\begin{displaymath}
\sum_{j}(\bar{\lambda}^{j\dagger})_{\alpha\beta}(\bar{\lambda}^{j})_{\gamma\delta} = \delta_{\alpha\delta}\delta_{\beta\gamma}-\frac{1}{2}\delta_{\alpha\beta}\delta_{\gamma\delta}\,,
\end{displaymath}
so that, for arbitrary $2\times2\,$ matrices $M,N$,\footnote{Further identities of this sort are $\sum_{j}\langle\bar{\lambda}^{j\dagger}\bar{\lambda}^{j}M\rangle=\frac{3}{2}\langle M\rangle$, $\sum_{j}\langle\bar{\lambda}^{j\dagger}M\bar{\lambda}^{j}\rangle=\frac{3}{2}\langle M\rangle$, or in general $\sum_{j}\langle\bar{\lambda}^{j\dagger}M\bar{\lambda}^{j}N\rangle=\langle M\rangle\langle N\rangle - \frac{1}{2}\langle MN\rangle$\,, summing over $j=\pi^{0},\pi^{+},\pi^{-}\,$. The anticommutator is $\lbrace\bar{\lambda}^{j},\,\bar{\lambda}^{k\dagger}\rbrace=\delta^{jk}\mathds{1}$.}:
\begin{displaymath}
\sum_{j}\langle\bar{\lambda}^{j\dagger}M\rangle\langle\bar{\lambda}^{j}N\rangle = \langle MN\rangle -\frac{1}{2}\langle M\rangle\langle N\rangle\,.
\end{displaymath}
We could equally well work directly with the Pauli matrices instead of the channel matrices. However, we prefer the channel matrix formalism here, since the generalization to $N>2$ flavors or to the case of isospin violation is straightforward therein, and since we are mostly interested in specific matrix elements involving the charged or neutral pions.\\
\underline{Vertices for $\mathbf{PP},\,\mathbf{SS}\,$:} \\
The (pseudo-)scalar external fields are $p=p^{0}\mathds{1}_{2\times 2}+p^{a}\tau^{a}$, $s=s^{0}\mathds{1}_{2\times 2}+s^{a}\tau^{a}$.
From $\mathcal{L}_{M}^{(2)}$ we can derive the vertex rules for $s^{i}\phi^2$ and $p^{i}\phi$ vertices. Expanding the Lagrangian density in $\phi$ and in the external fields $p,s$, we find
\begin{equation}\label{eq:L2Mexpanded}
\frac{F^{2}}{4}\langle\nabla_{\mu}U^{\dagger}\nabla^{\mu}U\rangle =  \frac{1}{2}\langle\partial_{\mu}\phi\partial^{\mu}\phi\rangle 
 + \frac{1}{12F^{2}}\langle\lbrack\phi,\partial_{\mu}\phi\rbrack\lbrack\phi,\partial^{\mu}\phi\rbrack\rangle\,+\ldots\,.
\end{equation}
From this we can construct a $\phi^{4}\,$- scattering vertex ($jl\rightarrow km$),
\begin{eqnarray}\label{eq:streu1}
 -\frac{i}{6F^{2}}&\biggl(&(p_{k}\cdot p_{m})(\langle\lbrack\bl^{j},\bl^{k\dagger}\rbrack\lbrack\bl^{l},\bl^{m\dagger}\rbrack\rangle + \langle\lbrack\bl^{j},\bl^{m\dagger}\rbrack\lbrack\bl^{l},\bl^{k\dagger}\rbrack\rangle) \nonumber \\ &-& (p_{k}\cdot p_{l})(\langle\lbrack\bl^{j},\bl^{k\dagger}\rbrack\lbrack\bl^{m\dagger},\bl^{l}\rbrack\rangle + \langle\lbrack\bl^{j},\bl^{l}\rbrack\lbrack\bl^{m\dagger},\bl^{k\dagger}\rbrack\rangle) \nonumber \\ &-& (p_{l}\cdot p_{m})(\langle\lbrack\bl^{j},\bl^{l}\rbrack\lbrack\bl^{k\dagger},\bl^{m\dagger}\rbrack\rangle + \langle\lbrack\bl^{j},\bl^{m\dagger}\rbrack\lbrack\bl^{k\dagger},\bl^{l}\rbrack\rangle) \nonumber \\ &-& (p_{j}\cdot p_{m})(\langle\lbrack\bl^{k\dagger},\bl^{j}\rbrack\lbrack\bl^{l},\bl^{m\dagger}\rbrack\rangle + \langle\lbrack\bl^{l},\bl^{j}\rbrack\lbrack\bl^{k\dagger},\bl^{m\dagger}\rbrack\rangle) \nonumber \\ &-& (p_{j}\cdot p_{k})(\langle\lbrack\bl^{l},\bl^{j}\rbrack\lbrack\bl^{m\dagger},\bl^{k\dagger}\rbrack\rangle + \langle\lbrack\bl^{m\dagger},\bl^{j}\rbrack\lbrack\bl^{l},\bl^{k\dagger}\rbrack\rangle) \nonumber \\ &+& (p_{j}\cdot p_{l})(\langle\lbrack\bl^{k\dagger},\bl^{j}\rbrack\lbrack\bl^{m\dagger},\bl^{l}\rbrack\rangle + \langle\lbrack\bl^{m\dagger},\bl^{j}\rbrack\lbrack\bl^{k\dagger},\bl^{l}\rbrack\rangle)\biggr)\,.
\end{eqnarray}
But also the ``mass term'' in $\mathcal{L}^{(2)}_{M}$ contains a $\phi^{4}\,$-contribution:
\begin{displaymath}
\frac{F^{2}}{4}\langle \chi U^{\dagger}+U\chi^{\dagger}\rangle = B\left(F^{2}\langle s\rangle + \sqrt{2}F\langle p\phi\rangle - \langle s\phi^{2}\rangle - \frac{\sqrt{2}}{3F}\langle p\phi^{3}\rangle + \frac{1}{6F^{2}}\langle(\mathcal{M}+\delta s)\phi^{4}\rangle + \ldots\right)\,.
\end{displaymath}
The vertex rule for a single pion coupling to the pseudoscalar source field is thus given by
\begin{equation}\label{eq:leadingpvertex}
 i\sqrt{2}BF\langle\bar{\lambda}^{j}\tau^{a}\rangle\,,
\end{equation}
for an incoming pion $j$. In particular, the pseudoscalar isosinglet source field does not couple to single pions (isovector particles) if isospin is conserved (note that the channel matrices are traceless).\\
The vertex rules for two pions coupling to the scalar field $s^{a}$ or $s^0$ are given by
\begin{equation}\label{eq:vertexspipi}
-iB\langle \tau^{a}\lbrace\bar{\lambda}^{j},\bar{\lambda}^{k\dagger}\rbrace\rangle\quad\mathrm{or}\quad -2iB\langle \bar{\lambda}^{j}\bar{\lambda}^{k\dagger}\rangle\,,
\end{equation}
for an incoming (outgoing) pion $j$ ($k$), while the vertices for three pions $j,k,l$ ($j,l$ in, $k$ out), coupling to $p^a$ are found as
\begin{equation}
-i\frac{\sqrt{2}B}{3F}\langle\tau^{a}\left(\bar{\lambda}^{j}\bar{\lambda}^{l}\bar{\lambda}^{k\dagger} + \mathrm{perm.\,of\,}\bar{\lambda}\right)\rangle\,,
\end{equation}
(for $p^0$, $\tau^{a}$ is replaced by $\mathds{1}_{2\times 2}$). Setting $m_{u}=m_{d}=:\hat{m}$, the scattering vertex from the second part of the Lagrangian is
\begin{eqnarray}\label{eq:streu2}
 & & i\frac{2B\hat{m}}{12F^{2}}\langle\bl^{j}\bl^{k\dagger}\bl^{l}\bl^{m\dagger} + \mathrm{perm.\,of\,}\bar{\lambda}\rangle \\ &=& i\frac{2B\hat{m}}{12F^{2}}\langle\lbrace\lbrace\bl^{j},\bl^{k\dagger}\rbrace,\lbrace\bl^{l},\bl^{m\dagger}\rbrace\rbrace + \lbrace\lbrace\bl^{j},\bl^{l}\rbrace,\lbrace\bl^{k\dagger},\bl^{m\dagger}\rbrace\rbrace + \lbrace\lbrace\bl^{j},\bl^{m\dagger}\rbrace,\lbrace\bl^{l},\bl^{k\dagger}\rbrace\rbrace\rangle\,.\nonumber
\end{eqnarray}
At leading order in the quark mass expansion, we can replace $2B\hat{m}\,\rightarrow\,M_{\pi}^{2}$.\\
\underline{Vertices for $\mathbf{AA},\,\mathbf{VV}\,$:} \\
The isovector source fields are taken as $v_{\mu}=v_{\mu}^{a}\frac{\tau^{a}}{2}$, $a_{\mu}=a_{\mu}^{a}\frac{\tau^{a}}{2}$. The necessary Feynman rules can be derived from the leading order chiral Lagrangian (\ref{eq:L2M}). In the expansion of the kinetic term, we now show the structures including one vector or axialvector source field:
\begin{eqnarray*}
\frac{F^{2}}{4}\langle\nabla_{\mu}U^{\dagger}\nabla^{\mu}U\rangle &=& \frac{F^{2}}{4}\langle u_{\mu}u^{\mu}\rangle = \frac{1}{2}\langle\partial_{\mu}\phi\partial^{\mu}\phi\rangle -\sqrt{2}F\langle(\partial_{\mu}\phi)a^{\mu}\rangle -\frac{\sqrt{2}}{3F}\langle\lbrack\phi,\partial_{\mu}\phi\rbrack\lbrack\phi,a^{\mu}\rbrack\rangle \\
 &+& i\langle(\partial_{\mu}\phi)\lbrack\phi,v^{\mu}\rbrack\rangle + \frac{1}{12F^{2}}\langle\lbrack\phi,\partial_{\mu}\phi\rbrack\lbrack\phi,\partial^{\mu}\phi\rbrack\rangle\,+\ldots
\end{eqnarray*}
There are also contact terms (``seagull terms'') involving two source fields, which result in contributions $\sim\delta^{4}(x-y)$ in position space. Being interested in large distances $x-y\not=0$, we do  not treat those terms here.
For an incoming pion $j$ with four-momentum $q^{\mu}$ coupling to a source field $a_{\mu}^{a}$, we read off the vertex factor
\begin{equation}
-q_{\mu}\frac{F}{\sqrt{2}}\langle\bar{\lambda}^{j}\tau^{a}\rangle\,.
\end{equation}
One also finds a $\phi^{3}a\,$ vertex ($jl\rightarrow k,a^{a}_{\mu}$)
\begin{eqnarray}
\frac{1}{3\sqrt{2}F}\biggl(&p_{k}^{\mu}&(\langle\lbrack \bl^{j},\bl^{k\dagger}\rbrack\lbrack \bl^{l},\tau^{a}\rbrack + \lbrack \bl^{l},\bl^{k\dagger}\rbrack\lbrack \bl^{j},\tau^{a}\rbrack\rangle) \label{eq:a3phivertex} \\ - &p_{j}^{\mu}&(\langle\lbrack \bl^{l},\bl^{j}\rbrack\lbrack \bl^{k\dagger},\tau^{a}\rbrack + \lbrack \bl^{k\dagger},\bl^{j}\rbrack\lbrack \bl^{l},\tau^{a}\rbrack\rangle) - p_{l}^{\mu}(\langle\lbrack \bl^{j},\bl^{l}\rbrack\lbrack \bl^{k\dagger},\tau^{a}\rbrack + \lbrack \bl^{k\dagger},\bl^{l}\rbrack\lbrack \bl^{j},\tau^{a}\rbrack\rangle)\biggr)\,,\nonumber
\end{eqnarray}
and the rule for a vector field $v_{\mu}^{a}$ attached to two pion fields $j\rightarrow k$,
\begin{equation}
\frac{i}{2}(p_{j}^{\mu}+p_{k}^{\mu})\langle\lbrack\bar{\lambda}^{j},\,\bar{\lambda}^{k\dagger}\rbrack\tau^{a}\rangle\,.
\end{equation}
\underline{Vertices involving spin 1 resonance fields:} \\
The leading Lagrangians for the interaction of the $\rho$ and the $a_{1}$ resonance fields with the pions are given in Eqs.~(\ref{eq:LV}) and (\ref{eq:LA}).
The building blocks occuring in these Lagrangians are collected in Eq.~(\ref{eq:resbuildingblocks}). We need the expansions
\begin{eqnarray}
u_{\mu} &=& 2a_{\mu}-\frac{\sqrt{2}}{F}\left(\partial_{\mu}\phi +i\lbrack \phi,\,v_{\mu}\rbrack\right) + \ldots\,,\\
F^{+}_{\mu\nu} &=& 2\left(\partial_{\mu}v_{\nu}-\partial_{\nu}v_{\mu}\right) - \frac{i\sqrt{2}}{F}\lbrack\phi,\,\partial_{\mu}a_{\nu}-\partial_{\nu}a_{\mu}\rbrack - \frac{1}{2F^{2}}\lbrack\phi,\,\lbrack\phi,\,\partial_{\mu}v_{\nu}-\partial_{\nu}v_{\mu}\rbrack\rbrack + \ldots\,,\label{eq:FplusExpanded}\\
\Gamma^{\mu} &=& -iv^{\mu} - \frac{1}{\sqrt{2}F}\lbrack\phi,\,a^{\mu}\rbrack + \frac{1}{4F^{2}}\lbrack\phi,\,\partial^{\mu}\phi\rbrack + \ldots\,,\label{eq:GammaExpanded}
\end{eqnarray}
\begin{equation}\label{eq:FminusExpanded}
-F^{-}_{\mu\nu} = 2\left(\partial_{\mu}a_{\nu}-\partial_{\nu}a_{\mu}\right) - \frac{i\sqrt{2}}{F}\lbrack\phi,\,\partial_{\mu}v_{\nu}-\partial_{\nu}v_{\mu}\rbrack - \frac{1}{2F^{2}}\lbrack\phi,\,\lbrack\phi,\,\partial_{\mu}a_{\nu}-\partial_{\nu}a_{\mu}\rbrack\rbrack + \ldots\,,
\end{equation}
from which one can obtain a vertex rule for $v_{\nu}^{b}(q)\rightarrow\rho_{\alpha}^{k}(q)$,
\begin{equation}\label{eq:vrhovertex}
-i\sqrt{2}f_{V}\left(q^2g^{\alpha\nu}-q^{\alpha}q^{\nu}\right)\langle\left(\frac{\tau^{b}}{2}\right)\bar{\lambda}^{k\dagger}\rangle\,,
\end{equation}
a rule for $\pi^{i}(p)a_{\nu}^{b}(q)\rightarrow\rho^{k}_{\beta}(l)$,
\begin{equation}\label{eq:apirhovertex}
-\left(\frac{2g_{V}}{F}\left((p\cdot l)g^{\nu\beta}-p^{\beta}l^{\nu}\right)+\frac{f_{V}}{F}\left((q\cdot l)g^{\nu\beta}-q^{\beta}l^{\nu}\right)\right)\langle\lbrack\bar{\lambda}^{i},\,\left(\frac{\tau^{b}}{2}\right)\rbrack\bar{\lambda}^{k\dagger}\rangle\,,
\end{equation}
 and also a vertex rule for $\pi^{i}(p)\rho^{k}_{\beta}(q)\rightarrow\pi^{j}(p')v_{\mu}^{a}(q')$,
\begin{eqnarray*}
 & & \frac{if_{V}}{2\sqrt{2}F^{2}}\left((q'\cdot q)g_{\beta\mu}-q'_{\beta}q_{\mu}\right)\left(\langle\lbrack\bar{\lambda}^{i},\,\lbrack\bar{\lambda}^{j\dagger},\,\left(\frac{\tau^{a}}{2}\right)\rbrack\rbrack\bar{\lambda}^{k}\rangle  +  \langle\lbrack\bar{\lambda}^{j\dagger},\,\lbrack\bar{\lambda}^{i},\,\left(\frac{\tau^{a}}{2}\right)\rbrack\rbrack\bar{\lambda}^{k}\rangle\right) \\
 &+& \frac{i\sqrt{2}g_{V}}{F^{2}}\left(\left((p'\cdot q)g_{\beta\mu}-p'_{\beta}q_{\mu}\right)\langle\lbrack\bar{\lambda}^{j\dagger},\,\lbrack\bar{\lambda}^{i},\,\left(\frac{\tau^{a}}{2}\right)\rbrack\rbrack\bar{\lambda}^{k}\rangle - \left((p\cdot q)g_{\beta\mu}-p_{\beta}q_{\mu}\right)\langle\lbrack\bar{\lambda}^{i},\,\lbrack\bar{\lambda}^{j\dagger},\,\left(\frac{\tau^{a}}{2}\right)\rbrack\rbrack\bar{\lambda}^{k}\rangle\right)\\
 &-& \biggl(\frac{if_{V}}{\sqrt{2}F^{2}}\left((\bar{p}\cdot q')g_{\beta\mu}-q'_{\beta}\bar{p}_{\mu}\right) + \frac{i\sqrt{2}g_{V}}{F^{2}}\left(p'_{\beta}p_{\mu}-p_{\beta}p'_{\mu}\right)\biggr)\langle\lbrack\bar{\lambda}^{i},\,\bar{\lambda}^{j\dagger}\rbrack\lbrack\left(\frac{\tau^{a}}{2}\right),\,\bar{\lambda}^{k}\rbrack\rangle\,. 
\end{eqnarray*}
\newpage
\underline{Vertices involving spin 0 resonance fields:} \\
The leading Lagrangians for the interaction of the $a_{0}$ and the $\eta$ resonance fields with the pions are given in Eqs.~(\ref{eq:LS})-(\ref{eq:Lsigma}).
Below we write out the necessary terms in the expansion of the building blocks:
\begin{eqnarray}
u_{\mu} &=& -\frac{\sqrt{2}}{F}\partial_{\mu}\phi + \ldots\,,\quad \Gamma_{\mu}=\frac{1}{4F^{2}}\lbrack\phi,\,\partial_{\mu}\phi\rbrack + \ldots\,,\\
\chi_{+} &=& 4Bs + \frac{2\sqrt{2}B}{F}\lbrace\phi,\,p\rbrace - \frac{B}{F^{2}}\lbrace\phi,\,\lbrace\phi,\,s\rbrace\rbrace - \frac{B}{3\sqrt{2}F^{3}}\lbrace\phi,\,\lbrace\phi,\,\lbrace\phi,\,p\rbrace\rbrace\rbrace  \nonumber \\ &+& \frac{B}{24F^{4}}\lbrace\phi,\,\lbrace\phi,\,\lbrace\phi,\,\lbrace\phi,\,s\rbrace\rbrace\rbrace\rbrace + \ldots\,,\label{eq:chiplusExpanded}\\
\chi_{-} &=& 4iBp - \frac{2\sqrt{2}iB}{F}\lbrace\phi,\,s\rbrace - \frac{iB}{F^{2}}\lbrace\phi,\,\lbrace\phi,\,p\rbrace\rbrace + \ldots\,.
\end{eqnarray}
The necessary vertex rules are easily read off from those expressions.

\newpage

\section{Momentum space integrals}
\label{app:integrale}
\def\theequation{\Alph{section}.\arabic{equation}}
\setcounter{equation}{0}

The master formula of dimensional regularization is (see e.~g. \cite{Collins:1984xc})
\begin{equation}\label{eq:A1}
\int\frac{d^d l_{\mathrm{Eucl.}}}{(2\pi)^d}\frac{(l^2_{\mathrm{Eucl.}})^{\alpha}}{(l^2_{\mathrm{Eucl.}}+M^2)^{\beta}} = \frac{M^{d+2\alpha -2\beta}}{(4\pi)^{\frac{d}{2}}}\frac{\Gamma(\beta-\alpha - \frac{d}{2})\Gamma(\frac{d}{2}+\alpha)}{\Gamma(\beta)\Gamma(\frac{d}{2})}\,.
\end{equation}
(Here $l_{\mathrm{Eucl.}}$ is the ``euclidean'' energy-momentum four-vector, with $l^{0}\equiv l^{0}_{\mathrm{Mink.}}= il^{0}_{\mathrm{Eucl.}}$.\\
For a space-like distance $x-y$ one can achieve $x^0=y^0$ by a convenient choice of the reference frame; due to lorentz-invariance, the integral
\begin{equation}\label{eq:defIM}
I_{M}(x-y):=\int\frac{d^d l}{(2\pi)^d}\frac{ie^{-il(x-y)}}{(l^2-M^{2})}
\end{equation}
does not depend on this choice. For $x\rightarrow y$, the integral diverges according to (\ref{eq:A1}) (for $d\rightarrow 4$),
\begin{equation}\label{eq:IM0}
I_{M}(0)=\frac{M^{d-2}}{(4\pi)^{\frac{d}{2}}}\Gamma\left(1-\frac{d}{2}\right)\,.
\end{equation}
For $x\not= y$ (which we will assume throughout), $x^0=y^0$ the $l^0$-integration can be performed in the standard fashion, employing a Wick-rotation and the residue theorem ($M^2\rightarrow M^2-i\epsilon$, $d\rightarrow 4$ etc.), and the integration over the space-like components is elementary,
\begin{eqnarray}
I_{M}(x-y) &=&\int\frac{d^3\mathbf{l}}{(2\pi)^3}\frac{e^{i\mathbf{l}\cdot(\mathbf{x}-\mathbf{y})}}{2\sqrt{\mathbf{l}^2+M^{2}}} \nonumber \\ 
 &=& \frac{1}{4\pi^2\left|\mathbf{x}-\mathbf{y}\right|}\int_{0}^{\infty}\frac{\left|\mathbf{l}\right|d\left|\mathbf{l}\right|}{\sqrt{\left|\mathbf{l}\right|^2 + M^2}}\sin\left(\left|\mathbf{l}\right|\left|\mathbf{x}-\mathbf{y}\right|\right) \nonumber \\ &=& \frac{M}{4\pi^2\left|\mathbf{x}-\mathbf{y}\right|}K_{1}(M\left|\mathbf{x}-\mathbf{y}\right|)\,,\label{eq:IM}
\end{eqnarray}
using the well-known formula
\begin{equation}\label{eq:BesselK}
\int_{0}^{\infty}\frac{u\,du}{\sqrt{u^2 +1}^{2n +1}}\sin uz = \frac{\sqrt{\pi}}{\Gamma(n+\frac{1}{2})}\left(\frac{z}{2}\right)^n K_{1-n}(z)\,,
\end{equation}
where $K_{\nu}(z)$ are the modified Bessel functions of the second kind (see e.~g.~\cite{WW:1996}), which obey
\begin{eqnarray}
\frac{2\nu}{z}K_{\nu}(z) &=& K_{\nu +1}(z)-K_{\nu -1}(z)\,,\label{eq:BesselKa}\\
\frac{dK_{\nu}(z)}{dz} &=& -\frac{1}{2}\left(K_{\nu -1}(z)+K_{\nu +1}(z)\right)\,.\label{eq:BesselKb}
\end{eqnarray}
In a similar way one finds
\begin{equation}\label{eq:IMM}
I_{MM}(x-y):=\int\frac{d^4 l}{(2\pi)^4}\frac{ie^{-il(x-y)}}{(l^2-M^{2})^2} = -\frac{1}{8\pi^2}K_{0}(M\left|\mathbf{x}-\mathbf{y}\right|) \overset{\mathrm{cov.}}{=}-\frac{1}{8\pi^2}K_{0}(M\sqrt{-(x-y)^2}) \,.
\end{equation}
From Lorentz invariance, we must have
\begin{equation}
\int\frac{d^4 l}{(2\pi)^4}\frac{il^{\mu}e^{-il(x-y)}}{(l^2-M^{2})} = (x-y)^{\mu}I_{M}^{(1)}\,,\quad \int\frac{d^4 l}{(2\pi)^4}\frac{il^{\mu}e^{-il(x-y)}}{(l^2-M^{2})^2} = (x-y)^{\mu}I_{MM}^{(1)}\,.
\end{equation}
For the scalar coefficients, one finds
\begin{eqnarray}
(x-y)^2 I_{MM}^{(1)} &=& \int\frac{d^4 l}{(2\pi)^4}\frac{il(x-y)e^{-il(x-y)}}{(l^2-M^{2})^2} = \frac{iM\left|\mathbf{x}-\mathbf{y}\right|}{8\pi^2}K_{1}(M\left|\mathbf{x}-\mathbf{y}\right|)\nonumber \\
\Rightarrow\,I_{MM}^{(1)} &=& \frac{1}{2i}I_{M}(x-y)\,,\label{eq:intlxmy}\\
I_{M}^{(1)} &=& \frac{iM^2}{4\pi^2\left|\mathbf{x}-\mathbf{y}\right|^2}K_{2}(M\left|\mathbf{x}-\mathbf{y}\right|)\,.\label{eq:intlxmy2}
\end{eqnarray}
Furthermore it is easy to see that
\begin{equation}\label{eq:lsqrint1}
\int\frac{d^4 l}{(2\pi)^4}\frac{il^{2}e^{-il(x-y)}}{(l^2-M^{2})^2} = I_{M}(x-y) + M^2I_{MM}(x-y) \,.
\end{equation}
Introducing a four-vector $\Delta$ and an integration over a Feynman-parameter $\alpha$, one also obtains
\begin{eqnarray}
I_{MM}^{\Delta}(x-y) &:=& \int\frac{d^4 l}{(2\pi)^4}\frac{ie^{-il(x-y)}}{((l-\frac{\Delta}{2})^2-M^2)((l+\frac{\Delta}{2})^2-M^2)} \nonumber \\ 
 &=& -\frac{1}{8\pi^2}\int_{-\frac{1}{2}}^{\frac{1}{2}}d\alpha\,e^{i\alpha\Delta(x-y)}K_{0}\left(M(\alpha)\left|\mathbf{x}-\mathbf{y}\right|\right)\,,
\end{eqnarray}
with
\begin{equation}
M(\alpha) = \sqrt{M^2 - \frac{\Delta^2}{4} + \Delta^2\alpha^2}\,.
\end{equation}
$M(\alpha)$ is real for $\Delta^2<4M^2$. A short calculation yields
\begin{equation}
\int\frac{d^4 l}{(2\pi)^4}\frac{i(l\cdot\Delta)e^{-il(x-y)}}{((l-\frac{\Delta}{2})^2-M^2)((l+\frac{\Delta}{2})^2-M^2)} = -i\sin\left(\frac{\Delta}{2}(x-y)\right)I_{M}(x-y)\,.
\end{equation}
In analogy to Eq.~(\ref{eq:intlxmy}), one finds
\begin{eqnarray}
iJ_{MM}^{\Delta} &:=& \int\frac{d^4 l}{(2\pi)^4}\frac{i(l\cdot(x-y))e^{-il(x-y)}}{((l-\frac{\Delta}{2})^2-M^2)((l+\frac{\Delta}{2})^2-M^2)}\nonumber \\
 &=& \frac{i}{8\pi^2}\int_{-\frac{1}{2}}^{\frac{1}{2}}d\alpha\,\biggl(M(\alpha)\left|\mathbf{x}-\mathbf{y}\right|\cos\left(\alpha\Delta(x-y)\right)K_{1}\left(M(\alpha)\left|\mathbf{x}-\mathbf{y}\right|\right) \nonumber \\ &+&  \alpha\Delta(x-y)\sin\left(\alpha\Delta(x-y)\right)K_{0}\left(M(\alpha)\left|\mathbf{x}-\mathbf{y}\right|\right)\biggr)\,,
\end{eqnarray}
so that we can also compute the following integral with a four-vector structure in the numerator,
\begin{equation}
 \int\frac{d^4 l}{(2\pi)^4}\frac{il^{\mu}e^{-il(x-y)}}{((l-\frac{\Delta}{2})^2-M^2)((l+\frac{\Delta}{2})^2-M^2)} = (x-y)^{\mu}I_{MM}^{\Delta(1)}(x-y) + \Delta^{\mu}I_{MM}^{\Delta(2)}(x-y)\,,
\end{equation}
with coefficient functions
\begin{eqnarray}
I_{MM}^{\Delta(1)}(x-y) &=& \frac{i}{k(\Delta)}\left(\Delta^2 J_{MM}^{\Delta} + (\Delta\cdot(x-y))\sin\left(\frac{\Delta}{2}(x-y)\right)I_{M}(x-y)\right),\\
I_{MM}^{\Delta(2)}(x-y) &=& \frac{-i}{k(\Delta)}\left((\Delta\cdot(x-y))J_{MM}^{\Delta} + (x-y)^2\sin\left(\frac{\Delta}{2}(x-y)\right)I_{M}(x-y)\right),\\
k(\Delta) &=& \Delta^2(x-y)^2 - (\Delta\cdot(x-y))^2\,.
\end{eqnarray}
For $\Delta\rightarrow 0$ we find $I_{MM}^{\Delta(1)}(x-y)\rightarrow I_{MM}^{(1)}(x-y)$ and $I_{MM}^{\Delta(2)}(x-y)\rightarrow 0$.
A generalization of (\ref{eq:lsqrint1}) is given by
\begin{equation}\label{eq:lsqrint2}
\int\frac{d^4l}{(2\pi)^4}\frac{il^2e^{-il(x-y)}}{((l-\frac{\Delta}{2})^2-M^2)((l+\frac{\Delta}{2})^2-M^2)} = \cos\left(\frac{\Delta}{2}(x-y)\right)I_{M}(x-y) + \left(M^2-\frac{\Delta^2}{4}\right)I_{MM}^{\Delta}(x-y)\,.
\end{equation}
We continue with tensor structures in the numerator: The integral
\begin{equation}\label{eq:lmunudef}
I_{MM}^{\mu\nu}:=\int\frac{d^4l}{(2\pi)^4}\frac{il^{\mu}l^{\nu}e^{-il(x-y)}}{((l-\frac{\Delta}{2})^2-M^2)((l+\frac{\Delta}{2})^2-M^2)}
\end{equation}
can be decomposed in the following way:
\begin{equation}
I_{MM}^{\mu\nu} = g^{\mu\nu}I_{MM}^{\Delta(3)} + (x-y)^{\mu}(x-y)^{\nu}I_{MM}^{\Delta(4)} + \Delta^{\mu}\Delta^{\nu}I_{MM}^{\Delta(5)} + \left(\Delta^{\mu}(x-y)^{\nu}+\Delta^{\nu}(x-y)^{\mu}\right)I_{MM}^{\Delta(6)}\,,
\end{equation}
\begin{eqnarray}
I_{MM}^{\Delta(3)} &=& \int_{-\frac{1}{2}}^{\frac{1}{2}}d\alpha\,\cos\left(\alpha\Delta(x-y)\right)\frac{M(\alpha)K_{1}\left(M(\alpha)\left|\mathbf{x}-\mathbf{y}\right|\right)}{8\pi^2\left|\mathbf{x}-\mathbf{y}\right|}\,,\\
I_{MM}^{\Delta(4)} &=& \int_{-\frac{1}{2}}^{\frac{1}{2}}d\alpha\,\cos\left(\alpha\Delta(x-y)\right)\frac{(M(\alpha))^2}{8\pi^2}\frac{K_{2}\left(M(\alpha)\left|\mathbf{x}-\mathbf{y}\right|\right)}{\left|\mathbf{x}-\mathbf{y}\right|^2}\,,\\
I_{MM}^{\Delta(5)} &=& -\int_{-\frac{1}{2}}^{\frac{1}{2}}d\alpha\,\cos\left(\alpha\Delta(x-y)\right)\frac{\alpha^2}{8\pi^2}K_{0}\left(M(\alpha)\left|\mathbf{x}-\mathbf{y}\right|\right)\,,\\
I_{MM}^{\Delta(6)} &=& -\int_{-\frac{1}{2}}^{\frac{1}{2}}d\alpha\,\sin\left(\alpha\Delta(x-y)\right)\frac{\alpha M(\alpha)K_{1}\left(M(\alpha)\left|\mathbf{x}-\mathbf{y}\right|\right)}{8\pi^2\left|\mathbf{x}-\mathbf{y}\right|}\,.
\end{eqnarray}
For $\Delta\rightarrow 0$, we find $I_{MM}^{\Delta(3)}\rightarrow\frac{1}{2}I_{M}(x-y)$, $I_{MM}^{\Delta(4)}\rightarrow\frac{1}{2i}I_{M}^{(1)}(x-y)$, $I_{MM}^{\Delta(5)}\rightarrow\frac{1}{12}I_{MM}(x-y)$ and $I_{MM}^{\Delta(6)}\rightarrow 0\,$.
For a simple propagator, the integral with tensor structure is
\begin{equation}\label{eq:IMtensordef}
I_{M}^{\mu\nu}:= \int\frac{d^4 l}{(2\pi)^4}\frac{il^{\mu}l^{\nu}e^{-il(x-y)}}{(l^2-M^{2})} = g^{\mu\nu}I_{M}^{(2)} + (x-y)^{\mu}(x-y)^{\nu}I_{M}^{(3)}\,,
\end{equation}
with coefficients
\begin{equation}\label{eq:IMtensorCoeff}
I_{M}^{(2)} = iI_{M}^{(1)}\,,\qquad I_{M}^{(3)} = - \frac{M^3}{4\pi^2}\frac{K_{3}\left(M\left|\mathbf{x}-\mathbf{y}\right|\right)}{\left|\mathbf{x}-\mathbf{y}\right|^3}\,.
\end{equation}
\begin{eqnarray}
 & & \int\frac{d^4l}{(2\pi)^4}\frac{il^{\mu}l^{\nu}l^2e^{-il(x-y)}}{((l-\frac{\Delta}{2})^2-M^2)((l+\frac{\Delta}{2})^2-M^2)} = \cos\left(\frac{\Delta}{2}(x-y)\right)\left(I_{M}^{\mu\nu}+\frac{\Delta^{\mu}\Delta^{\nu}}{4}I_{M}\right) \nonumber \\ &-& \frac{i}{2}\sin\left(\frac{\Delta}{2}(x-y)\right)\left(\Delta^{\mu}(x-y)^{\nu}+\Delta^{\nu}(x-y)^{\mu}\right)I_{M}^{(1)} + \left(M^2-\frac{\Delta^2}{4}\right)I_{MM}^{\mu\nu}\,.
\end{eqnarray}
We also need to consider the case with three open Lorentz indices,
\begin{eqnarray}
I_{MM}^{\mu\nu\rho} &:=& \int\frac{d^4l}{(2\pi)^4}\frac{il^{\mu}l^{\nu}l^{\rho}e^{-il(x-y)}}{((l-\frac{\Delta}{2})^2-M^2)((l+\frac{\Delta}{2})^2-M^2)}\nonumber \\
 &=& \left(g^{\mu\nu}(x-y)^{\rho}+g^{\mu\rho}(x-y)^{\nu}+g^{\nu\rho}(x-y)^{\mu}\right)I_{MM}^{\Delta(7)} + (x-y)^{\mu}(x-y)^{\nu}(x-y)^{\rho}I_{MM}^{\Delta(8)}  \nonumber \\ &+& \left(g^{\mu\nu}\Delta^{\rho}+g^{\mu\rho}\Delta^{\nu}+g^{\nu\rho}\Delta^{\mu}\right)I_{MM}^{\Delta(9)} +\Delta^{\mu}\Delta^{\nu}\Delta^{\rho}I_{MM}^{\Delta(10)} \nonumber \\ &+& \left((x-y)^{\mu}(x-y)^{\nu}\Delta^{\rho} + (x-y)^{\mu}(x-y)^{\rho}\Delta^{\nu} + (x-y)^{\nu}(x-y)^{\rho}\Delta^{\mu}\right)I_{MM}^{\Delta(11)}  \nonumber \\
 &+& \left((x-y)^{\mu}\Delta^{\nu}\Delta^{\rho} + (x-y)^{\nu}\Delta^{\rho}\Delta^{\mu} + (x-y)^{\rho}\Delta^{\mu}\Delta^{\nu}\right)I_{MM}^{\Delta(12)} \,,\label{eq:lmunurhodecomp}
\end{eqnarray}
\begin{eqnarray}
I_{MM}^{\Delta(7)} &=& iI_{MM}^{\Delta(4)}\,,\nonumber \\
I_{MM}^{\Delta(8)} &=& i\int_{-\frac{1}{2}}^{\frac{1}{2}}d\alpha\,\cos\left(\alpha\Delta(x-y)\right)\frac{(M(\alpha))^3}{8\pi^2}\frac{K_{3}\left(M(\alpha)\left|\mathbf{x}-\mathbf{y}\right|\right)}{\left|\mathbf{x}-\mathbf{y}\right|^3}\,,\quad I_{MM}^{\Delta(9)} = iI_{MM}^{\Delta(6)}\,,\nonumber \\
I_{MM}^{\Delta(10)} &=& i\int_{-\frac{1}{2}}^{\frac{1}{2}}d\alpha\,\sin\left(\alpha\Delta(x-y)\right)\frac{\alpha^3}{8\pi^2}K_{0}\left(M(\alpha)\left|\mathbf{x}-\mathbf{y}\right|\right)\,,\nonumber \\
I_{MM}^{\Delta(11)} &=& -i\int_{-\frac{1}{2}}^{\frac{1}{2}}d\alpha\,\sin\left(\alpha\Delta(x-y)\right)\frac{\alpha(M(\alpha))^2}{8\pi^2}\frac{K_{2}\left(M(\alpha)\left|\mathbf{x}-\mathbf{y}\right|\right)}{\left|\mathbf{x}-\mathbf{y}\right|^2}\,,\nonumber \\
I_{MM}^{\Delta(12)} &=& -i\int_{-\frac{1}{2}}^{\frac{1}{2}}d\alpha\,\cos\left(\alpha\Delta(x-y)\right)\frac{\alpha^2M(\alpha)K_{1}\left(M(\alpha)\left|\mathbf{x}-\mathbf{y}\right|\right)}{8\pi^2\left|\mathbf{x}-\mathbf{y}\right|}\,.
\end{eqnarray}
For $\Delta\rightarrow 0$, only the coefficients $I_{MM}^{\Delta(7,8)}$ remain, with $I_{MM}^{\Delta(7)}\rightarrow\frac{1}{2}I_{M}^{(1)}(x-y)$ and 
\begin{equation}
I_{MM}^{\Delta(8)}\overset{\Delta\rightarrow 0}{\longrightarrow} \frac{iM^3}{8\pi^2}\frac{K_{3}\left(M\left|\mathbf{x}-\mathbf{y}\right|\right)}{\left|\mathbf{x}-\mathbf{y}\right|^3} = \frac{1}{2i}I_{M}^{(3)}\,.
\end{equation}
In reference frames where $x^0\not=y^0$, one has to substitute $\left|\mathbf{x}-\mathbf{y}\right|\rightarrow\sqrt{-(x-y)^2}$ everywhere.

\newpage

\section{Derivation of soft-pion theorems}
\label{app:softpions}
\def\theequation{\Alph{section}.\arabic{equation}}
\setcounter{equation}{0}

In this appendix, we employ the method of soft-pion theorems (see \cite{Nambu:1997wa,Adler:1964um,Adler:1965ga,Weinberg:1966fm,Dashen:1969ez,deAlfaro:1973zz,Weinberg:1967kj}, or Sec.~11.3 of \cite{Itzykson:1980rh} for a textbook treatment), 
to demonstrate some general relations between limiting values of matrix elements studied in this work.
To this end, we neglect the pion mass and treat only the chiral limit, so that the isovector axial-vector current is also conserved (in addition to the vector current).
We assume that the matrix elements considered here do not diverge in the chiral limit.
For two arbitrary local operators $O_{1,2}(x)$ we then consider the following fourier-transformed matrix element:
\begin{equation}\label{eq:defMOO}
M_{\rho\sigma}(x,y,p',p) := \int d^{4}z'\int d^{4}z\,e^{ip'z'}e^{-ipz}\langle 0|TA_{\rho}^{j}(z')A_{\sigma}^{i}(z)O_{1}(x)O_{2}(y)|0\rangle\,,
\end{equation}
and compute
\begin{eqnarray*}
p'^{\rho}p^{\sigma}M_{\rho\sigma} &=& \int d^{4}z'\int d^{4}z\,\left(\partial^{\rho}_{z'}e^{ip'z'}\right)\left(\partial^{\sigma}_{z}e^{-ipz}\right)\langle 0|TA_{\rho}^{j}(z')A_{\sigma}^{i}(z)O_{1}(x)O_{2}(y)|0\rangle \\ &=& \int d^{4}z'\int d^{4}z\,e^{ip'z'}e^{-ipz}\partial^{\rho}_{z'}\partial^{\sigma}_{z}\langle 0|TA_{\rho}^{j}(z')A_{\sigma}^{i}(z)O_{1}(x)O_{2}(y)|0\rangle \\
 &=& \int d^{4}z'\int d^{4}z\,e^{ip'z'}e^{-ipz}\partial^{\rho}_{z'}(\langle 0|T\lbrack A_{0}^{i}(z),A_{\rho}^{j}(z')\rbrack O_{1}(x)O_{2}(y)|0\rangle\delta(z^{0}-z'^{0}) \\ & &  \hspace{4.3cm} +\, \langle 0|TA_{\rho}^{j}(z')\lbrack A_{0}^{i}(z),O_{1}(x)\rbrack O_{2}(y)|0\rangle\delta(z^{0}-x^{0}) \\ & & \hspace{4.3cm} +\, \langle 0|TA_{\rho}^{j}(z')O_{1}(x)\lbrack A_{0}^{i}(z),O_{2}(y)\rbrack|0\rangle\delta(z^{0}-y^{0}))\,.
\end{eqnarray*}
The commutators stem from the derivatives of the theta functions which are implicit in the time-ordering operation.
In order to continue, we need the commutators (summation over the isospin index $k$ implied, see e.~g. \cite{Scherer:2002tk}, Sec.~2.4, for the explicit calculation)
\begin{eqnarray*}
\lbrack A_{0}^{i}(z),A_{\rho}^{j}(z')\rbrack_{z^{0}=z'^{0}} &=& i\epsilon^{ijk}V^{k}_{\rho}(z)\delta^{3}(\vec{z}-\vec{z}\,')\,,\quad \lbrack A_{0}^{i}(z),V_{\rho}^{j}(z')\rbrack_{z^{0}=z'^{0}} = i\epsilon^{ijk}A^{k}_{\rho}(z)\delta^{3}(\vec{z}-\vec{z}\,')\,,\\
\lbrack V_{0}^{i}(z),A_{\rho}^{j}(z')\rbrack_{z^{0}=z'^{0}} &=& i\epsilon^{ijk}A^{k}_{\rho}(z)\delta^{3}(\vec{z}-\vec{z}\,')\,,\quad \lbrack V_{0}^{i}(z),V_{\rho}^{j}(z')\rbrack_{z^{0}=z'^{0}} = i\epsilon^{ijk}V^{k}_{\rho}(z)\delta^{3}(\vec{z}-\vec{z}\,')\,.
\end{eqnarray*}
Using these commutation relations one obtains
\begin{eqnarray*}
p'^{\rho}p^{\sigma}M_{\rho\sigma} &=& \int d^{4}z'\int d^{4}z\,e^{ip'z'}e^{-ipz}\biggl(\langle 0|T\lbrack\lbrack A_{0}^{i}(z),A_{0}^{j}(z')\rbrack,O_{1}(x)\rbrack O_{2}(y)|0\rangle\delta(z^{0}-z'^{0})\delta(z'^{0}-x^{0}) \\ & & \hspace{4cm} + \,\langle 0|TO_{1}(x)\lbrack\lbrack A_{0}^{i}(z),A_{0}^{j}(z')\rbrack,O_{2}(y)\rbrack|0\rangle\delta(z^{0}-z'^{0})\delta(z'^{0}-y^{0}) \\
 & & \hspace{4cm} + \,\langle 0|T\lbrack A_{0}^{j}(z'),\lbrack A_{0}^{i}(z),O_{1}(x)\rbrack\rbrack O_{2}(y)|0\rangle\delta(z^{0}-x^{0})\delta(z'^{0}-x^{0}) \\
 & & \hspace{4cm} + \,\langle 0|T\lbrack A_{0}^{i}(z),O_{1}(x)\rbrack\lbrack A_{0}^{j}(z'),O_{2}(y)\rbrack|0\rangle\delta(z^{0}-x^{0})\delta(z'^{0}-y^{0}) \\
 & & \hspace{4cm} + \,\langle 0|T\lbrack A_{0}^{j}(z'),O_{1}(x)\rbrack\lbrack A_{0}^{i}(z),O_{2}(y)\rbrack|0\rangle\delta(z^{0}-y^{0})\delta(z'^{0}-x^{0}) \\
 & & \hspace{4cm} + \,\langle 0|TO_{1}(x)\lbrack A_{0}^{j}(z'),\lbrack A_{0}^{i}(z),O_{2}(y)\rbrack\rbrack|0\rangle\delta(z^{0}-y^{0})\delta(z'^{0}-y^{0})\biggr)\,.
\end{eqnarray*}
Considering the case where $p'$ und $p$ approach zero, and defining
\begin{equation}
Q_{A}^{i}(t) := \int d^{3}\vec{z}\,A^{i}_{0}(\vec{z},t)\,,
\end{equation}
we evaluate all integrations involving delta functions and arrive at an interesting result:
\begin{eqnarray*}
p'^{0}p^{0}M_{00} &\overset{p^{0},p'^{0}\rightarrow 0}{\longrightarrow}& \langle 0|T\lbrack Q^{i}_{A}(x^{0}),\lbrack Q^{j}_{A}(x^{0}),O_{1}(x)\rbrack\rbrack O_{2}(y)|0\rangle + \langle 0|TO_{1}(x)\lbrack Q^{i}_{A}(y^{0}),\lbrack Q^{j}_{A}(y^{0}),O_{2}(y)\rbrack\rbrack|0\rangle \\ &+& \langle 0|T\lbrack Q^{i}_{A}(x^{0}),O_{1}(x)\rbrack\lbrack Q^{j}_{A}(y^{0}),O_{2}(y)\rbrack|0\rangle + \langle 0|T\lbrack Q^{j}_{A}(x^{0}),O_{1}(x)\rbrack\lbrack Q^{i}_{A}(y^{0}),O_{2}(y)\rbrack|0\rangle\,.
\end{eqnarray*}
If and only if the r.h.s. does {\em not}\, vanish, $M_{00}(x,y,p',p)$ must contain a term with a double pole $\sim\frac{1}{p'^{0}}\frac{1}{p^{0}}$. 
Such contributions can exist because $\langle 0|A_{\mu}|\pi\rangle$ does not vanish, so that the operators $A$ can generate a (massless) pion.
Therefore we can extract the pole contribution in question ($\grave a$ {\em la} LSZ \cite{Lehmann:1954rq}, using translation invariance):
\begin{equation}
M_{00}(x,y,p',p)|_{\vec{p}\,'\rightarrow 0,\vec{p}\rightarrow 0} = \langle 0|A^{j}_{0}(0)|\pi^{j}\rangle\frac{i}{(p'^{0})^2}\langle\pi^{j}|TO_{1}(x)O_{2}(y)|\pi^{i}\rangle\frac{i}{(p^{0})^2}\langle\pi^{i}|A^{i}_{0}(0)|0\rangle
+ \ldots\,,
\end{equation}
(where the dots stand for terms without the double pole) and conclude, with $\langle 0|A^{j}_{0}(0)|\pi^{j}(p')\rangle = ip'^{0}F_{\pi}$ and the result above,
\begin{eqnarray}
\langle\pi^{j}(p')|TO_{1}(x)O_{2}(y)|\pi^{i}(p)\rangle \overset{p,p'\rightarrow 0}{\longrightarrow} &-& \frac{1}{F_{\pi}^{2}}\biggl(\langle 0|T\lbrack Q^{i}_{A}(x^{0}),\lbrack Q^{j}_{A}(x^{0}),O_{1}(x)\rbrack\rbrack O_{2}(y)|0\rangle \nonumber \\ &+& \langle 0|TO_{1}(x)\lbrack Q^{i}_{A}(y^{0}),\lbrack Q^{j}_{A}(y^{0}),O_{2}(y)\rbrack\rbrack|0\rangle \nonumber \\ &+& \langle 0|T\lbrack Q^{i}_{A}(x^{0}),O_{1}(x)\rbrack\lbrack Q^{j}_{A}(y^{0}),O_{2}(y)\rbrack|0\rangle \nonumber \\ &+& \langle 0|T\lbrack Q^{j}_{A}(x^{0}),O_{1}(x)\rbrack\lbrack Q^{i}_{A}(y^{0}),O_{2}(y)\rbrack|0\rangle\biggr)\,.\label{eq:softpi}
\end{eqnarray}
In this way the (zero-momentum) pion matrix elements can be expressed through vacuum expectation values. A simple corollary (setting $O_{2}=\mathds{1}$) is
\begin{equation}
\langle\pi^{j}(p')|O_{1}(x)|\pi^{i}(p)\rangle \overset{p,p'\rightarrow 0}{\longrightarrow} -\frac{1}{F_{\pi}^{2}}\langle 0|T\lbrack Q^{i}_{A}(x^{0}),\lbrack Q^{j}_{A}(x^{0}),O_{1}(x)\rbrack\rbrack|0\rangle\,.
\end{equation}
Here is an example: For $O_{1}=S^{0}=\bar{q}{q}$ and
\begin{eqnarray}
\lbrack Q^{i}_{A}(x^{0}),S^{0}(x)\rbrack &=& iP^{i}(x)\,,\quad \lbrack Q^{i}_{A}(x^{0}),P^{j}(x)\rbrack =  -i\delta^{ij}S^{0}(x)\,,\\
\lbrack Q^{i}_{A}(x^{0}),S^{j}(x)\rbrack &=& i\delta^{ij}P^{0}(x)\,,\quad \lbrack Q^{i}_{A}(x^{0}),P^{0}(x)\rbrack =  -iS^{i}(x)
\end{eqnarray}
one finds (setting $F_{\pi}\rightarrow F$, the decay constant in the chiral limit) 
\begin{equation}
 \langle\pi^{j}(p')|S^{0}(x)|\pi^{i}(p)\rangle \overset{p,p'\rightarrow 0}{\longrightarrow} -\frac{\delta^{ij}}{F^{2}}\langle 0|S^{0}(x)|0\rangle = +\frac{2BF^{2}}{F^{2}}\delta^{ij}=2B\delta^{ij}\,,
\end{equation}
see Eq.~(\ref{eq:scalff}) or Eqs.~(11.3), (15.9) in \cite{Gasser:1983yg}. A more interesting result can be derived from the commutators
\begin{equation}
\lbrack Q^{i}_{A}(x^{0}),V^{a}_{\mu}(x)\rbrack = i\epsilon^{iak}A^{k}_{\mu}(x)\,,\quad \lbrack Q^{i}_{A}(x^{0}),A^{a}_{\mu}(x)\rbrack = i\epsilon^{iak}V^{k}_{\mu}(x)\,,
\end{equation}
which follow directly from the commutation relations stated above. One finds
\begin{displaymath}
\lbrack Q^{i}_{A},\lbrack Q^{j}_{A},A^{a}_{\mu}\rbrack\rbrack = \delta^{ij}A^{a}_{\mu}-\delta^{ia}A^{j}_{\mu}\,,\quad \lbrack Q^{i}_{A},\lbrack Q^{j}_{A},V^{a}_{\mu}\rbrack\rbrack = \delta^{ij}V^{a}_{\mu}-\delta^{ia}V^{j}_{\mu}\,,
\end{displaymath}
and using our theorem (\ref{eq:softpi}) together with the definitions
\begin{equation}\label{eq:defcAV}
\langle 0|TA^{a}_{\mu}(x)A^{b}_{\nu}(y)|0\rangle = \delta^{ab}c^{A}_{\mu\nu}(x-y)\,,\quad \langle 0|TV^{a}_{\mu}(x)V^{b}_{\nu}(y)|0\rangle = \delta^{ab}c^{V}_{\mu\nu}(x-y)\,,
\end{equation}
we arrive at
\begin{eqnarray}
\langle\pi^{j}(p')|TA^{a}_{\mu}(x)A^{b}_{\nu}(y)|\pi^{i}(p)\rangle &\rightarrow& -\frac{1}{F^{2}}\left(2\delta^{ij}\delta^{ab}-\delta^{ia}\delta^{jb}-\delta^{ib}\delta^{ja}\right)\left(c^{A}_{\mu\nu}(x-y)-c^{V}_{\mu\nu}(x-y)\right), \nonumber \\ \label{eq:TAApipi} \\
\langle\pi^{j}(p')|TV^{a}_{\mu}(x)V^{b}_{\nu}(y)|\pi^{i}(p)\rangle &\rightarrow& -\frac{1}{F^{2}}\left(2\delta^{ij}\delta^{ab}-\delta^{ia}\delta^{jb}-\delta^{ib}\delta^{ja}\right)\left(c^{V}_{\mu\nu}(x-y)-c^{A}_{\mu\nu}(x-y)\right),\nonumber \\ \label{eq:TVVpipi}
\end{eqnarray}
and in particular
\begin{equation}\label{eq:VV+AA=0}
\langle\pi^{j}(p')|T(A^{a}_{\mu}(x)A^{b}_{\nu}(y)+V^{a}_{\mu}(x)V^{b}_{\nu}(y))|\pi^{i}(p)\rangle \rightarrow 0\,.
\end{equation}
This result is valid in the chiral limit and for vanishing pion momenta, but for an arbitrary distance $x-y$: in this respect, it goes beyond the ChPT results.\\
It is reassuring to see that our results for the correlators are in agreement with (\ref{eq:VV+AA=0}). Comparing with the representation in \cite{Gasser:1983yg} (Eqs.~(12.1) and (13.2) in this reference), on tree level, and in the limit $M_{\pi}\rightarrow 0$, $p,p'\rightarrow 0$, 
\begin{equation}
c^{A}_{GL,\mu\nu}(x-y) = F_{\pi}^{2}\int\frac{d^{4}q}{(2\pi)^{4}}e^{-iq(x-y)}\frac{iq_{\mu}q_{\nu}}{q^{2}+i\epsilon}+\ldots,\quad c^{V}_{GL,\mu\nu}(x-y) = 0+\ldots,
\end{equation}
leaving out higher-order terms and a contact-term contribution $\sim\delta^{4}(x-y)$ (``Seagull''-graph), with our results for $\langle\pi|VV|\pi\rangle$ in the same limit (see Eqs.~(\ref{eq:resC1VV})-(\ref{eq:resC3VV}))
\begin{eqnarray}
\langle\pi|VV|\pi\rangle &\rightarrow& + \frac{1}{4}\int\frac{d^{4}q}{(2\pi)^{4}}e^{-iq(x-y)}\frac{iq_{\mu}q_{\nu}}{q^{2}+i\epsilon}\left(\langle\lbrack\tau^{a},\bar{\lambda}^{j\dagger}\rbrack\lbrack\bar{\lambda}^{i},\tau^{b}\rbrack\rangle + \langle\lbrack\tau^{b},\bar{\lambda}^{j\dagger}\rbrack\lbrack\bar{\lambda}^{i},\tau^{a}\rbrack\rangle\right) \nonumber \\
 &=& +\left(2\delta^{ij}\delta^{ab}-\delta^{ia}\delta^{jb}-\delta^{ib}\delta^{ja}\right)\int\frac{d^{4}q}{(2\pi)^{4}}e^{-iq(x-y)}\frac{iq_{\mu}q_{\nu}}{q^{2}+i\epsilon}\,,
\end{eqnarray}
we find consistency with Eq.~(\ref{eq:TVVpipi}) (due to (\ref{eq:VV+AA=0}) this also holds for the corresponding limit of $\langle\pi|AA|\pi\rangle$). - Further examples:
\begin{eqnarray}
\langle\pi^{j}(p')|TS^{0}(x)S^{0}(y)|\pi^{i}(p)\rangle &\rightarrow& -\frac{1}{F^{2}}2\delta^{ij}\left(\tilde{c}^{S}(x-y)-c^{P}(x-y)\right)\,, \label{eq:TS0S0pipi} \\
\langle\pi^{j}(p')|TP^{0}(x)P^{0}(y)|\pi^{i}(p)\rangle &\rightarrow& -\frac{1}{F^{2}}2\delta^{ij}\left(\tilde{c}^{P}(x-y)-c^{S}(x-y)\right)\,, \label{eq:TP0P0pipi} \\
\langle\pi^{j}(p')|TS^{a}(x)S^{b}(y)|\pi^{i}(p)\rangle &\rightarrow& -\frac{1}{F^{2}}\left(\delta^{ia}\delta^{jb}+\delta^{ib}\delta^{ja}\right)\left(c^{S}(x-y)-\tilde{c}^{P}(x-y)\right)\,, \label{eq:TSSpipi} \\
\langle\pi^{j}(p')|TP^{a}(x)P^{b}(y)|\pi^{i}(p)\rangle &\rightarrow& -\frac{1}{F^{2}}\left(\delta^{ia}\delta^{jb}+\delta^{ib}\delta^{ja}\right)\left(c^{P}(x-y)-\tilde{c}^{S}(x-y)\right)\,, \label{eq:TPPpipi} 
\end{eqnarray}
where we have used similar definitions as in (\ref{eq:defcAV}),
\begin{eqnarray}
\langle 0|TS^{0}(x)S^{0}(y)|0\rangle &=& \delta^{ij}\tilde{c}^{S}(x-y)\,,\quad \langle 0|TP^{0}(x)P^{0}(y)|0\rangle = \delta^{ij}\tilde{c}^{P}(x-y)\,,\\
\langle 0|TS^{a}(x)S^{b}(y)|0\rangle &=& \delta^{ab}c^{S}(x-y)\,,\quad \langle 0|TP^{a}(x)P^{b}(y)|0\rangle = \delta^{ab}c^{P}(x-y)\,.
\end{eqnarray}
In the representation of Gasser und Leutwyler \cite{Gasser:1983yg}, again on tree level (LO):
\begin{eqnarray*}
\tilde{c}^{S}_{GL}(x-y) &=& 0+\ldots\,,\quad \tilde{c}^{P}_{GL}(x-y) = 0+\ldots\,,\quad c^{S}_{GL}(x-y)=0+\ldots\,,\\
c^{P}_{GL}(x-y) &=& G_{\pi}^{2}\int\frac{d^{4}q}{(2\pi)^{4}}\frac{ie^{-iq(x-y)}}{q^{2}+i\epsilon}+\ldots\,,
\end{eqnarray*}
with $G_{\pi}\rightarrow 2BF$, see (\ref{eq:Gpi}). Our corresponding expressions for the pion matrix elements in the chiral limit and for $p,p'\rightarrow 0$ (see Eqs.~(\ref{eq:resC1PP})-(\ref{eq:resC3PP}),(\ref{eq:defIM}),(\ref{eq:IMM})) were
\begin{eqnarray}
\langle\pi^{j}(p')|TS^{0}(x)S^{0}(y)|\pi^{i}(p)\rangle &\rightarrow& 8B^{2}\delta^{ij}\int\frac{d^{4}q}{(2\pi)^{4}}\frac{ie^{-iq(x-y)}}{q^{2}+i\epsilon}+\ldots\,,\\
\langle\pi^{j}(p')|TP^{0}(x)P^{0}(y)|\pi^{i}(p)\rangle &\rightarrow& 0+\ldots\,,\\
\langle\pi^{j}(p')|TS^{a}(x)S^{b}(y)|\pi^{i}(p)\rangle &\rightarrow& 0+\ldots\,,\\
\langle\pi^{j}(p')|TP^{a}(x)P^{b}(y)|\pi^{i}(p)\rangle &\rightarrow& \left(\delta^{ia}\delta^{jb}+\delta^{ib}\delta^{ja}\right)\left(-4B^{2}I_{M}(x-y)|_{M_{\pi}\rightarrow 0}\right)+\ldots \nonumber \\ 
 &=& -4B^{2}\left(\delta^{ia}\delta^{jb}+\delta^{ib}\delta^{ja}\right)\int\frac{d^{4}q}{(2\pi)^{4}}\frac{ie^{-iq(x-y)}}{q^{2}+i\epsilon}+\ldots\,.
\end{eqnarray}
Again we find consistency with the theorems (\ref{eq:TS0S0pipi})-(\ref{eq:TPPpipi}). Note that $C_{1}\rightarrow 0$ in Eq.~(\ref{eq:resC1PP}), because $I_{MM}^{\Delta}\rightarrow I_{MM}\sim K_{0}(M_{\pi}r)$ diverges only as $\log(M_{\pi}r)$ for $M_{\pi}\rightarrow 0$, while $C_{3}$ in (\ref{eq:resC3PP}) vanishes for $\bar{p}\cdot(x-y)\rightarrow 0$. In general, for vanishing momenta,
\begin{equation}
I^{\Delta(1)}_{MM}\overset{\Delta\rightarrow 0}{\longrightarrow} I^{(1)}_{MM}\overset{(\ref{eq:intlxmy})}{=}\frac{1}{2i}I_{M}\overset{M\rightarrow 0}{\longrightarrow} -\frac{i}{8\pi^2|\mathbf{x}-\mathbf{y}|^2
}\,.
\end{equation}

\end{appendix}

\end{document}